\documentclass[11pt]{article}
\usepackage{xspace}
\usepackage{graphicx}
\usepackage{amsmath}
\usepackage{amssymb}
\usepackage{color}
\usepackage{url}
\usepackage{hyperref}
\usepackage{lineno}

\definecolor{Red}{rgb}{1,0,0}
\definecolor{Green}{rgb}{0,1,0}
\definecolor{Blue}{rgb}{0,0,1}
\definecolor{Black}{rgb}{0,0,0}

\def\beq{\begin{equation}}
\def\eeq#1{\label{#1}\end{equation}}
\def\eeqn{\end{equation}}

\def\beqa{\begin{eqnarray}}
\def\eeqa#1{\label{#1}\end{eqnarray}}
\def\eeqan{\end{eqnarray}}

\let\bar=\overbar

\def\Dslash{\not{\hbox{\kern-4pt $D$}}}
\def\dslash{\not{\hbox{\kern-2pt $\del$}}}

\def\msb{{\bar{\ssstyle M \kern -1pt S}}}

\newcommand{\parenbar}[2][4]{%
  \mkern#1mu
  \sbox0{$#2$}%
  \makebox[0pt][r]{\raisebox{\ht0}{$\scriptscriptstyle($}}%
  \overline{\mkern-#1mu#2\mkern-1mu}%
  \makebox[0pt][l]{\raisebox{\ht0}{$\scriptscriptstyle)$}}%
  \mkern1mu
}

\newcommand\ifb{fb$^{-1}$}
\newcommand\met{$E_{\text{T}}^{\text{miss}}$}
\newcommand\pt{$p_{\text{T}}$}
\def\Title#1{\begin{center} {\Large {\bf #1} } \end{center}}

\begin{document}
\Title{Exotics Searches at ATLAS}

\bigskip\bigskip


\begin{raggedright}  

Ruth P\"ottgen\index{P\"ottgen, R.}, {\it CERN/Johannes Gutenberg - University Mainz}\\

\begin{center}\emph{On behalf of the ATLAS Collaboration.}\end{center}
\bigskip
\end{raggedright}

{\small
\begin{flushleft}
\emph{To appear in the proceedings of the Interplay between Particle and Astroparticle Physics workshop, 18 -- 22 August, 2014, held at Queen Mary University of London, UK.}
\end{flushleft}
}

\section{Introduction}
The Standard Model (SM) of particle physics is one of the most successful theories in the history of science. To date it has been tested in numerous experiments with tremendous precision and so far withstands all experimental tests. However, there is a number of experimental phenomena that cannot be accommodated within the Standard Model. Among those are the hierarchy problem, the matter-antimatter asymmetry in the universe, as well as the existence of dark matter and dark energy. Various models for {\it physics beyond the Standard Model} (BSM) exist that provide solutions for one or the other of these open questions. Models for BSM physics (excluding super-symmetry) are collectively referred to as ``exotics" in the following. These proceedings will summarise some of the most recent results on searches for these models from the ATLAS experiment at the Large Hadron Collider~\cite{LHC} (LHC). The general strategy is very similar for all the searches reported on: The distribution of a discriminant variable is compared between data and Standard Model background in search for a deviation from the Standard Model prediction. In case no deviation is found, limits on model parameters are derived. The background is often estimated in a data-driven way to minimise the dependency on modelling in the simulation.

\section{ATLAS and the LHC}

During its first data taking run from spring 2010 to early 2013, the LHC has provided approximately 5\,fb$^{-1}$ of proton-proton collisions at a centre-of-mass energy of $\sqrt{s}=7$\,TeV and about 20\,fb$^{-1}$ at $\sqrt{s}=8$\,TeV. 
Since February 2013, the LHC is in a two-year shutdown phase to prepare for centre-of-mass energies of up to 14\,TeV and instantaneous luminosities beyond 10$^{34}$\,cm$^{-2}$s$^{-1}$.\\
ATLAS is a general purpose detector, designed to cover a broad physics programme and to detect and identify all products of the $pp$ collisions. It consists of three main components: the inner detector tracking system, the calorimeters and the muon system. The inner detector is immersed in a solenoidal magnetic field; the magnetic field in the muon system is generated by three large toroidal magnets. A detailed description of the ATLAS detector can be found in Ref.~\cite{ATLAS}.  
If not stated otherwise, the analyses presented here use the full data set of $\sqrt{s}=8$\,TeV $pp$ collisions. 

\section{Results from Various Searches} 
\subsection{General Search}
\begin{figure}[!ht]
\begin{center}
\includegraphics[width=0.5\columnwidth]{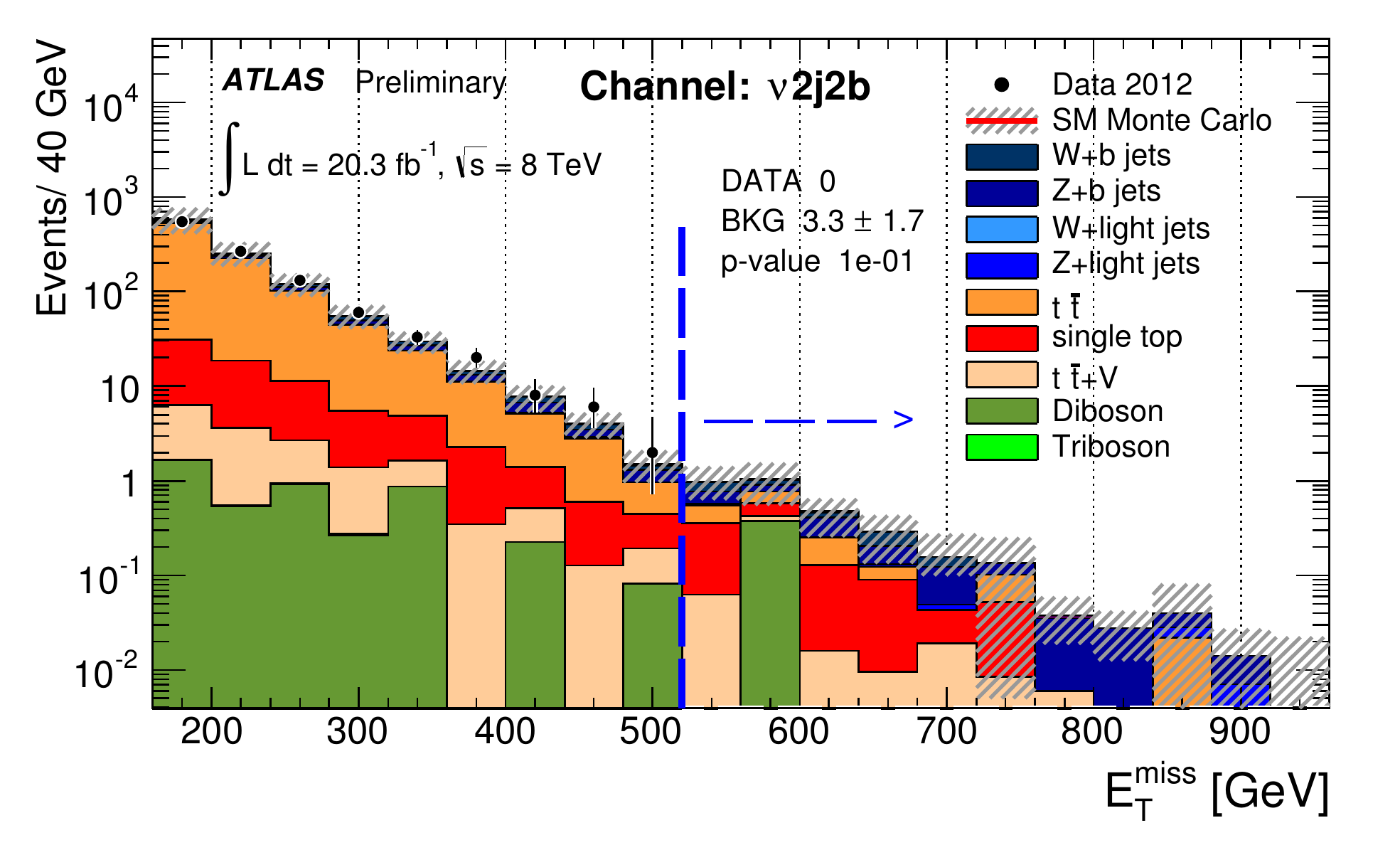}
\caption{\met{} distribution for events with two jets, two b-jets and missing transverse energy. Data (black points) is compared to the Standard Model background prediction (coloured histograms). The region of interest with the smallest $p$-value for this event class is indicated by the dashed line and arrow. Figure taken from Ref.~\cite{genSearch}}
\label{fig:genSearch}
\end{center}
\end{figure}

One possible approach to search for new physics is to not restrict the analysis to any specific model, but to look for deviations from the Standard Model in as many distributions as possible, as was done in Ref.~\cite{genSearch}. While such a {\it general search} is not as sensitive as an optimised specific search, it allows for a comprehensive investigation and has the virtue of providing the possibility to detect signals that are not predicted by any theory yet. The analysis uses approximately 700 event classes that are categorised according to the final states, which can involve electrons, photons, muons, jets, jets from b-quarks (b-jets) and missing transverse energy (\met{}). Three discriminant variables are used: the missing transverse energy, the scalar sum of the transverse momentum (\pt{}) of all objects including \met{}, referred to as {\it effective mass} or $m_{\text{eff}}$, and the invariant mass ($m_{\text{inv}}$) of all objects excluding \met{}. An algorithm searches for the regions of largest deviation in these three variables for all the event classes. An example is given in figure~\ref{fig:genSearch}: It shows the \met{} spectrum for events with missing transverse energy, two jets, and two b-jets. For \met{}$>520$\,GeV (indicated by the dashed line and arrow), the discrepancy between data and background model corresponds to a $p$-value of 0.1. Taking the look-elsewhere-effect into account, no significant deviation from the Standard Model is found in this analysis.
%
%
\subsection{New Heavy Quarks}
\begin{figure}[!ht]
\begin{center}
\includegraphics[width=0.35\columnwidth]{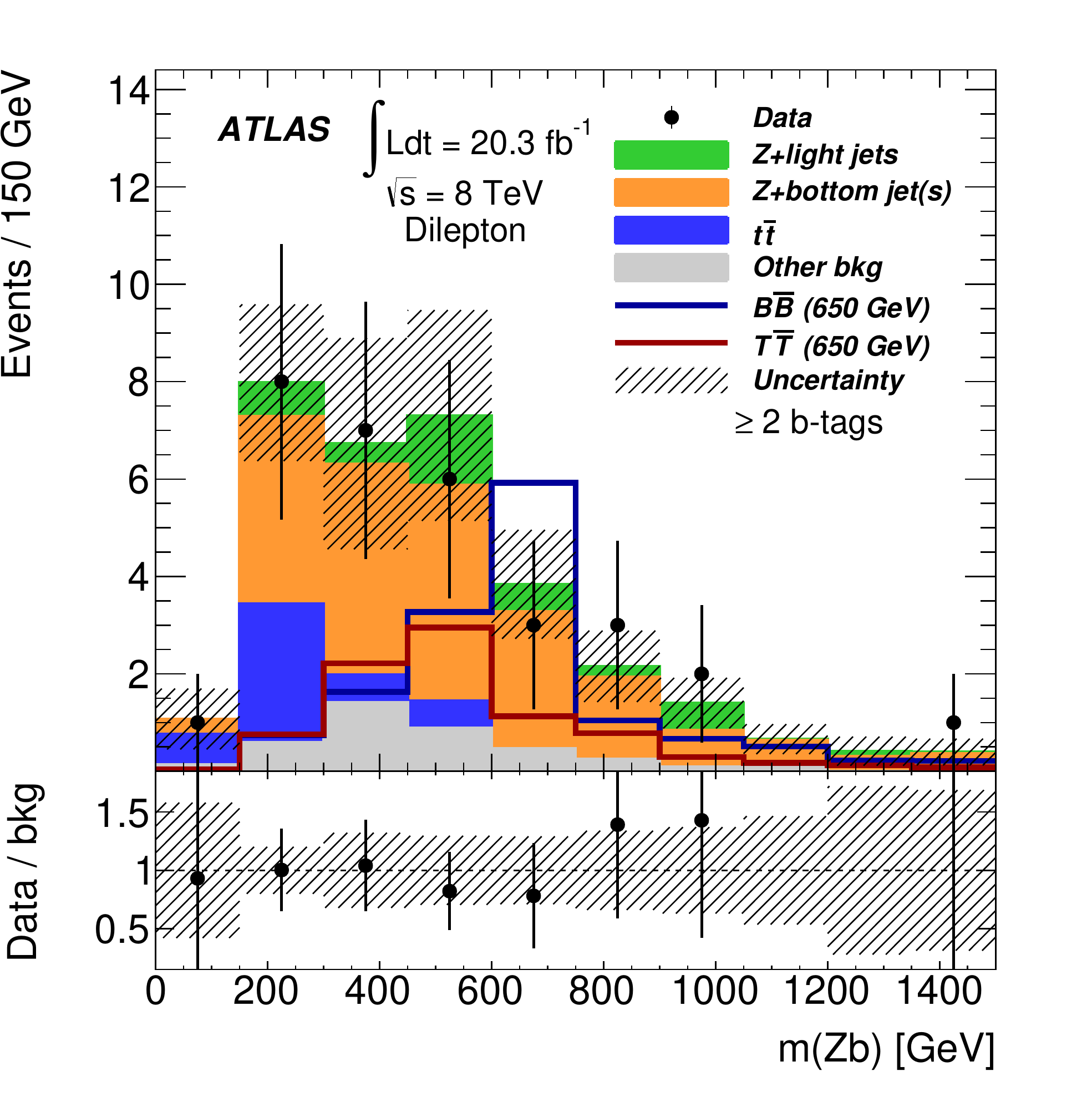}
\includegraphics[width=0.47\columnwidth]{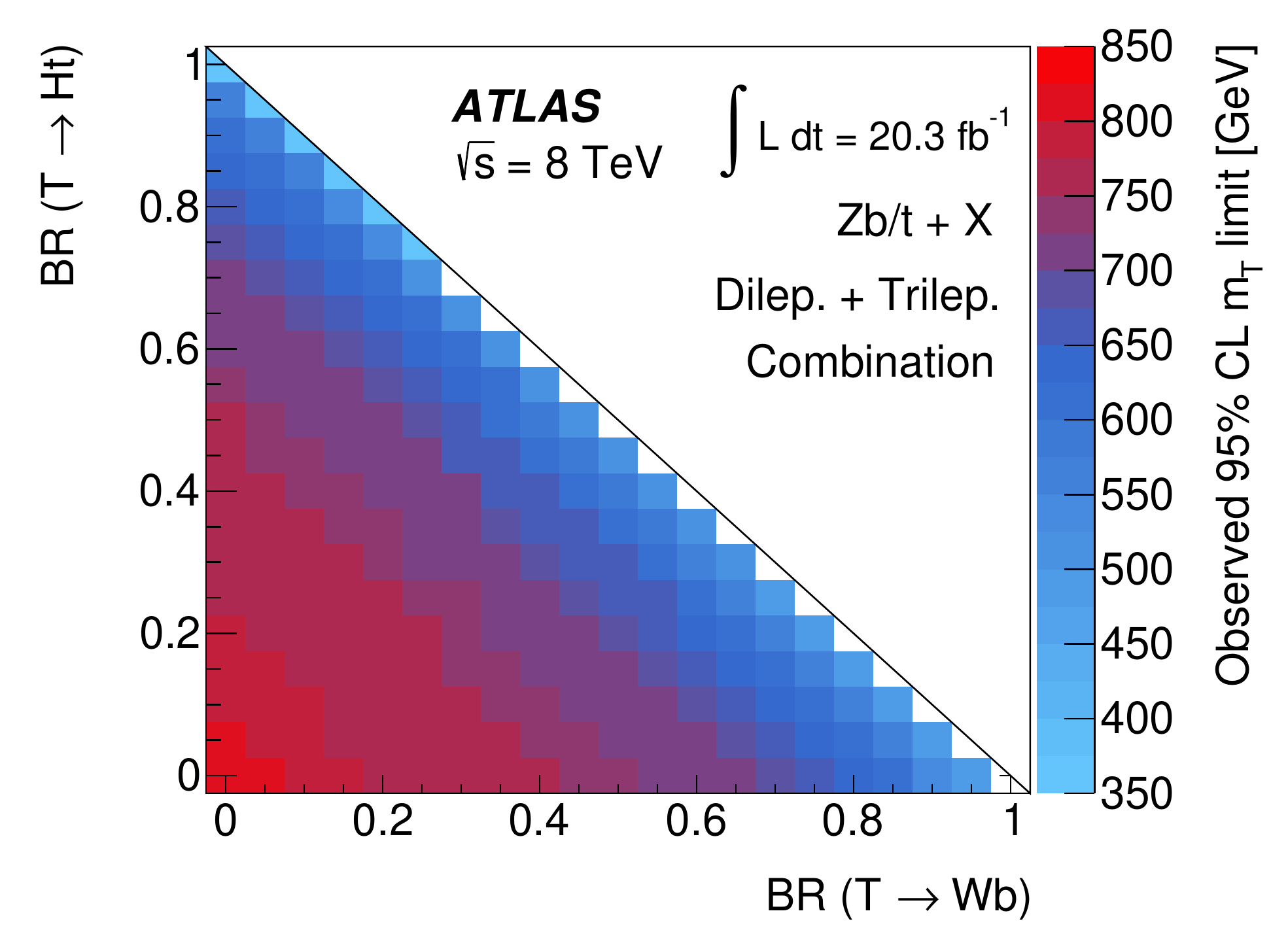}
\caption{Mass of the $Z$-$b$-system in the di-lepton channel (left) and combined observed lower 95\%CL limit on the mass of the $T$-vector-quark in the plane of the ($T\rightarrow Wb$, $T\rightarrow Ht$) branching ratio. Figures taken from Ref.~\cite{heavyQuarks}}
\label{fig:plotsHQ}
\end{center}
\end{figure}

Various non-SUSY natural models address the problem of divergences in the one-loop corrections to the Higgs mass by the introduction of vector-like quarks, see for example Ref.~\cite{hqTH} (for a more exhaustive list of model references please consult Ref.~\cite{heavyQuarks}). A search for singly and pair produced heavy vector quarks has been performed by the ATLAS collaboration~\cite{heavyQuarks}. The analysis searches for events with a heavy vector-like quark, labelled $T$ or $B$, decaying into a $Z$-boson and a SM $t$ or $b$ quark, respectively. One example is the pair production where at least one of the vector-like quarks decays to $Z$+SM-quark. 
The analysis selects events with a $Z$-boson with a transverse momentum greater than 150\,GeV, where the $Z$ is reconstructed from either two muons or two electrons. In addition, the presence of jets or b-jets is required and the analysis is performed in two categories of lepton multiplicity: either exactly two or more than two. The discriminant variables used are the scalar \pt{} sum of the jets and leptons ($H_{\text{T}}$) or the mass of the $Z$-$b$-jet system ($m(Zb)$). The distribution of the latter in the di-lepton category with two or more b-tagged jets is presented in the left panel of figure~\ref{fig:plotsHQ}. 
For illustration, also distributions for two signal points are included (solid lines). No significant deviation from the Standard Model is observed and limits on the mass of the new heavy quarks are derived for different assumptions on the branching ratios. The right panel of figure~\ref{fig:plotsHQ} shows the observed limit on the mass of the $T$ quark at 95\%CL for various configurations of branching ratios into $Wb$ and $Ht$. These limits are obtained by combining the two lepton multiplicity categories. The limits range from approximately 350\,GeV in the case of large branching ratio for one of the other channels to about 850\,GeV in case both branching ratios are close to zero.
%
\subsection{New Gauge Bosons}
Many scenarios for BSM physics predict the existence of new gauge bosons, for example models for grand unified theories (GUTs) or solutions to the hierarchy problem, see for example Ref.~\cite{Altarelli:1989ff,London:1986dk} and the more exhaustive list in Ref.~\cite{Wprime,Zprime}. These new gauge bosons are expected to decay similarly to the SM $W$ and $Z$ boson, and are thus searched for in events with a lepton and a neutrino (giving rise to missing transverse energy) or with two electrons or muons, respectively. 
\subsubsection{Lepton plus \met{}}
\label{subsubsec:lepMet}
\begin{figure}[!ht]
\begin{center}
\includegraphics[width=0.45\columnwidth]{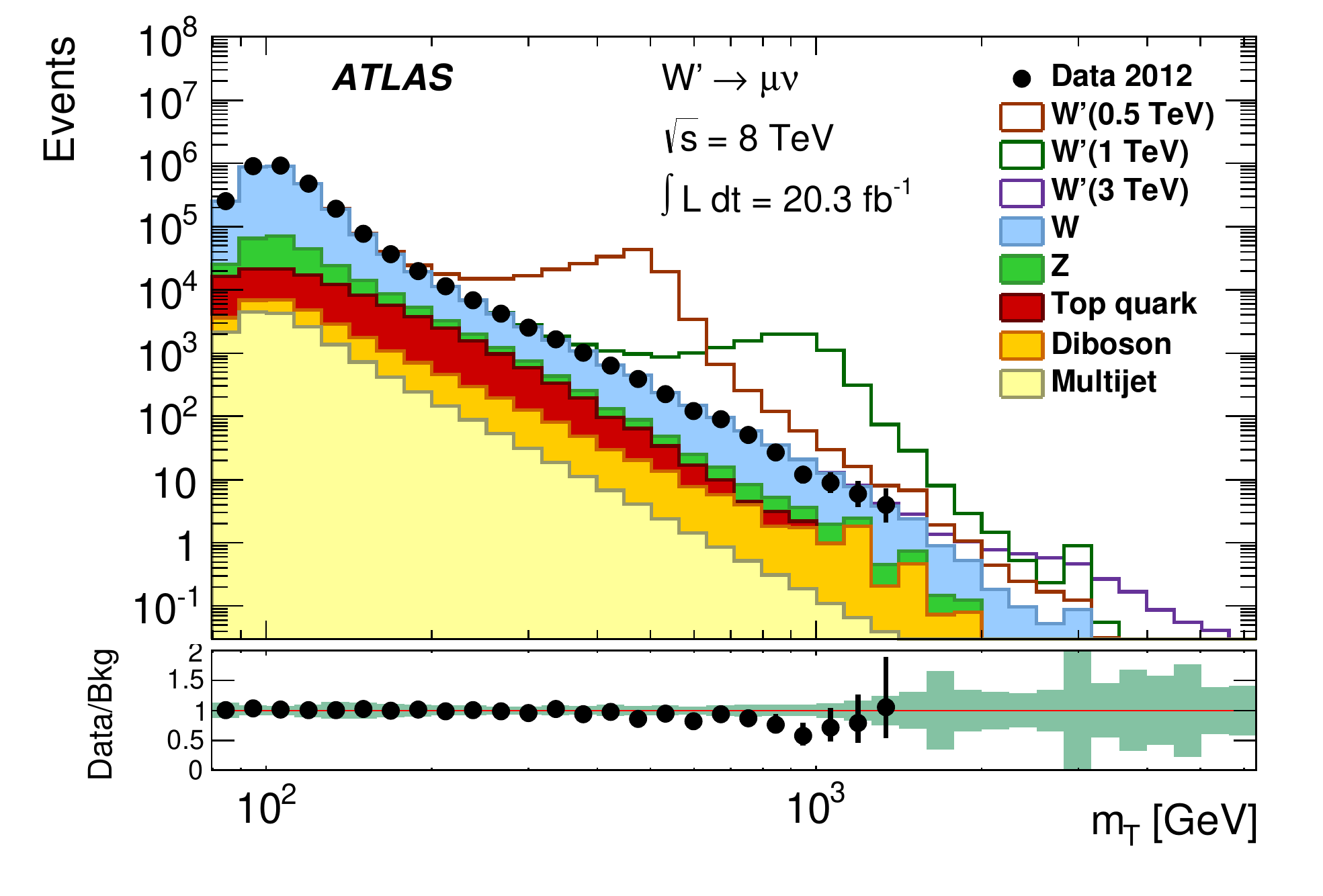}
\includegraphics[width=0.38\columnwidth]{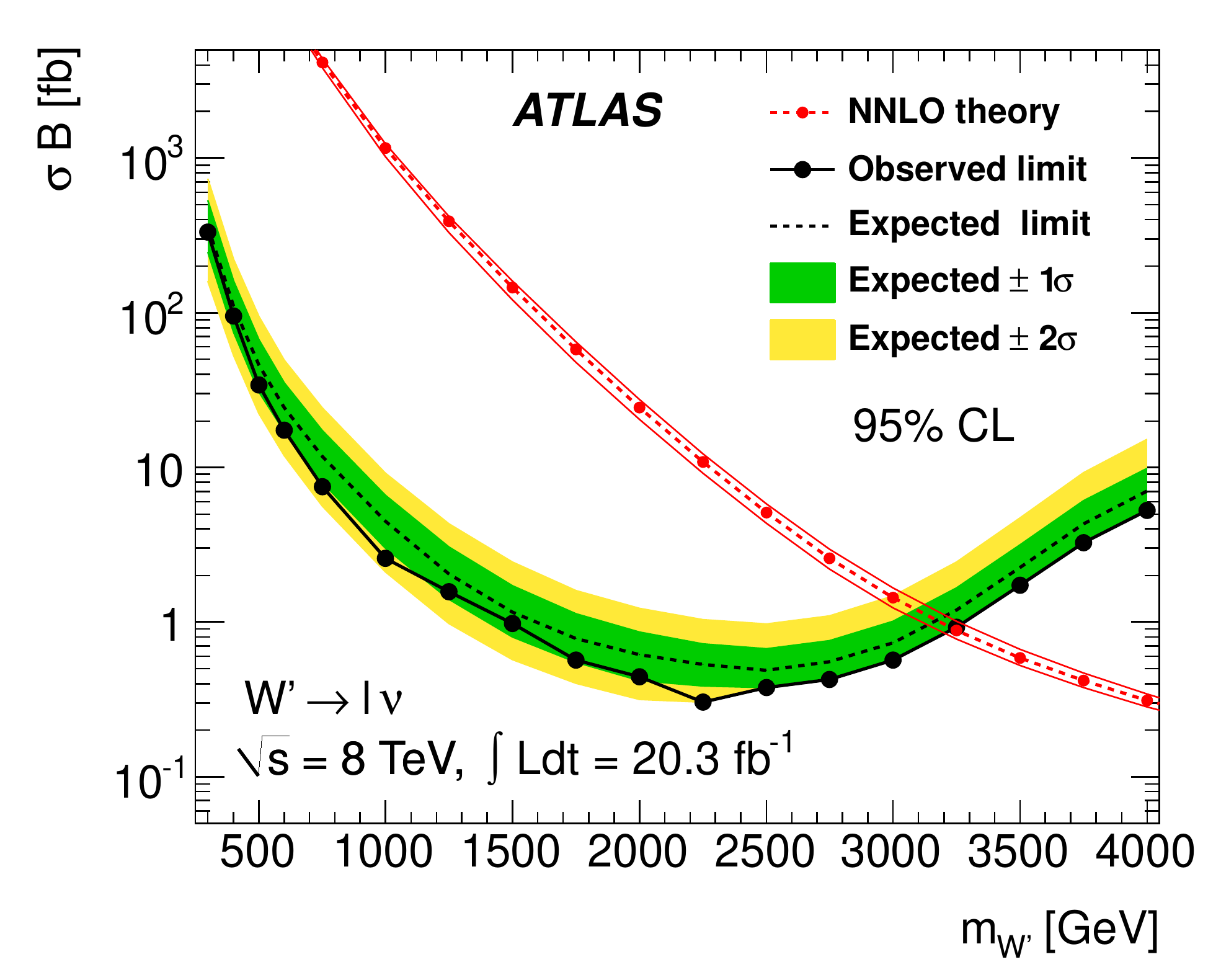}
\caption{Transverse mass of the muon-\met{}-system (left) and 95\%CL combined limit on the $W'$ mass (right). Figures taken from Ref.~\cite{Wprime}}
\label{fig:plotsWprime}
\end{center}
\end{figure}

In the lepton plus \met{} analysis~\cite{Wprime}, selected events are required to have exactly one electron (muon) fulfilling a certain set of quality criteria and with a \pt{} greater than 125\,GeV (45\,GeV). The same cut is placed on the \met{}. The discriminant variable is the transverse mass $m_{T}$\footnote{$m_{T}$ is calculated from the lepton \pt{}, the neutrino \pt{} (i.e. \met{}) and the angle between the two: $m_{\text{T}}=\sqrt{2p_{\text{T}}E_{\text{T}}^{\text{miss}}(1-\cos(\Delta\phi)_{\ell,\nu})}$}, the distribution of which is presented in the left panel of figure~\ref{fig:plotsWprime}. In addition to the data (black points) and the SM background (filled histograms), distributions for three signal points are displayed, showing that a possible signal would occur at the tail of the distribution. However, no such excess is observed, and limits are set on the mass of a heavy gauge boson for different models. One example of the 95\%CL limit on the cross section times branching ratio, as obtained from the combination of electron and muon channel, is shown in the right panel of figure~\ref{fig:plotsWprime}. The observed limit is given by a solid black line with points, the expected limit as a dashed black line, together with the $\pm1\sigma$ and $\pm2\sigma$ bands in green and yellow. The next-to-next-to-leading (NNLO) theory expectation is displayed in red. The observed lower limit on the mass of the $W'$ boson is obtained to be 3.24\,TeV.
%
\subsubsection{Di-Lepton}
\begin{figure}[!ht]
\begin{center}
\includegraphics[width=0.44\columnwidth]{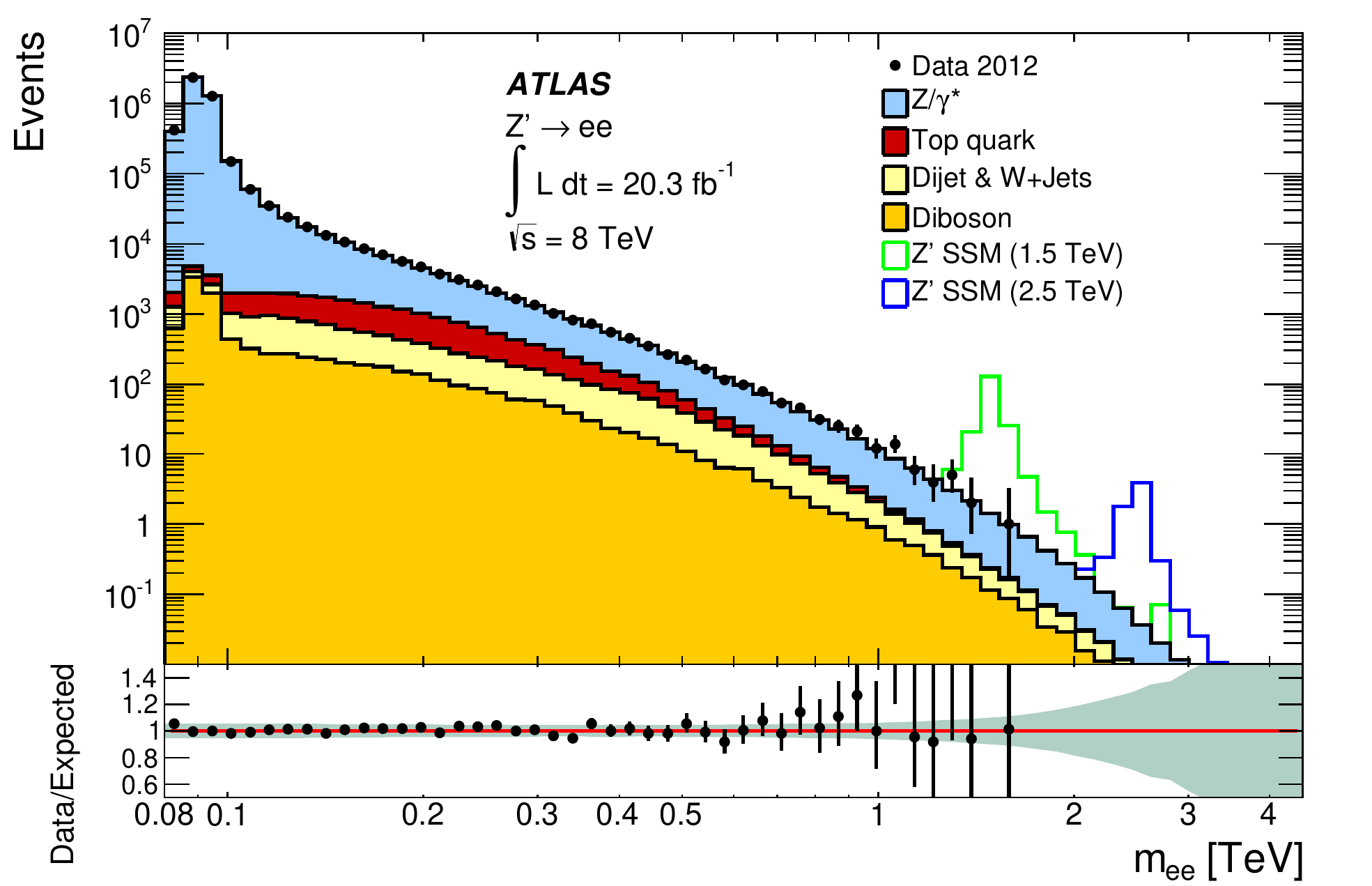}
\includegraphics[width=0.41\columnwidth]{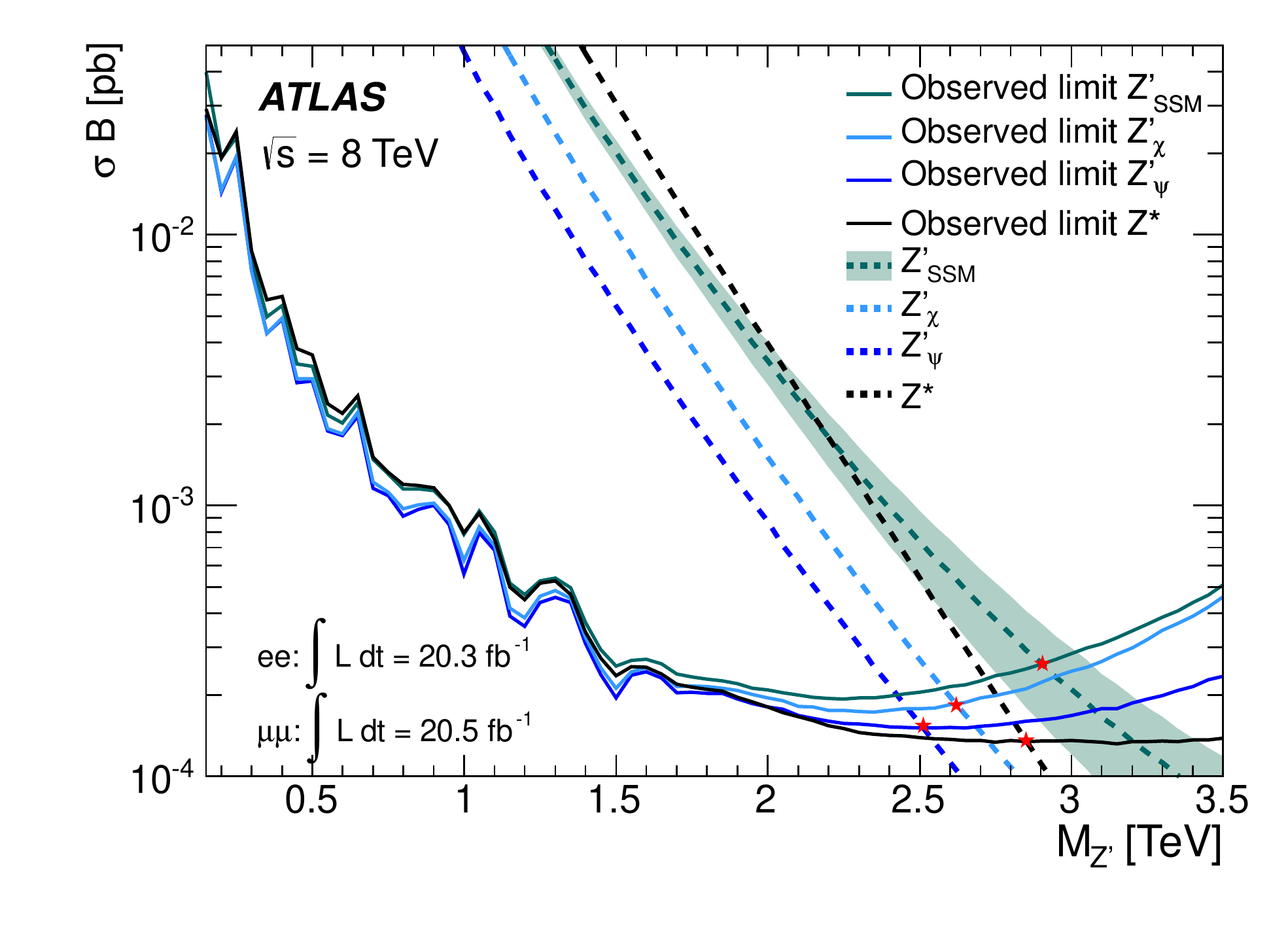}
\caption{Invariant mass of the di-electron system (left) and 95\%CL combined limit on the mass of the heavy gauge boson for different models (right). Figures taken from Ref.~\cite{Zprime}}
\label{fig:plotsZprime}
\end{center}
\end{figure}

The ATLAS collaboration also performed a search for resonances in the di-lepton invariant mass spectrum~\cite{Zprime}, both in the electron as well as in the muon channel. In the electron channel, the (sub)leading electron has to have \pt{}$>40(30)$\,GeV. In the muon channel, both muons are required to have \pt{}$>25$\,GeV and opposite charge. For both channels, the invariant mass is considered between 128\,GeV and 4.5\,TeV. The left panel of figure~\ref{fig:plotsZprime} shows the invariant mass distribution obtained in the electron channel, the peak at the $Z$-mass of approximately 90\,GeV is clearly visible. The data is found to be in very good agreement with the SM prediction. The green and blue histogram illustrate the signal expectations for two example points. The 95\%CL limits on the cross section times branching ratio for various scenarios are presented in the right panel of figure~\ref{fig:plotsZprime}, together with the theory expectation. The limits on the mass of the heavy gauge boson range from 2.46\,TeV to 2.87\,TeV.
%
\subsection{Contact Interaction}
\begin{figure}[!ht]
\begin{center}
\includegraphics[width=0.36\columnwidth]{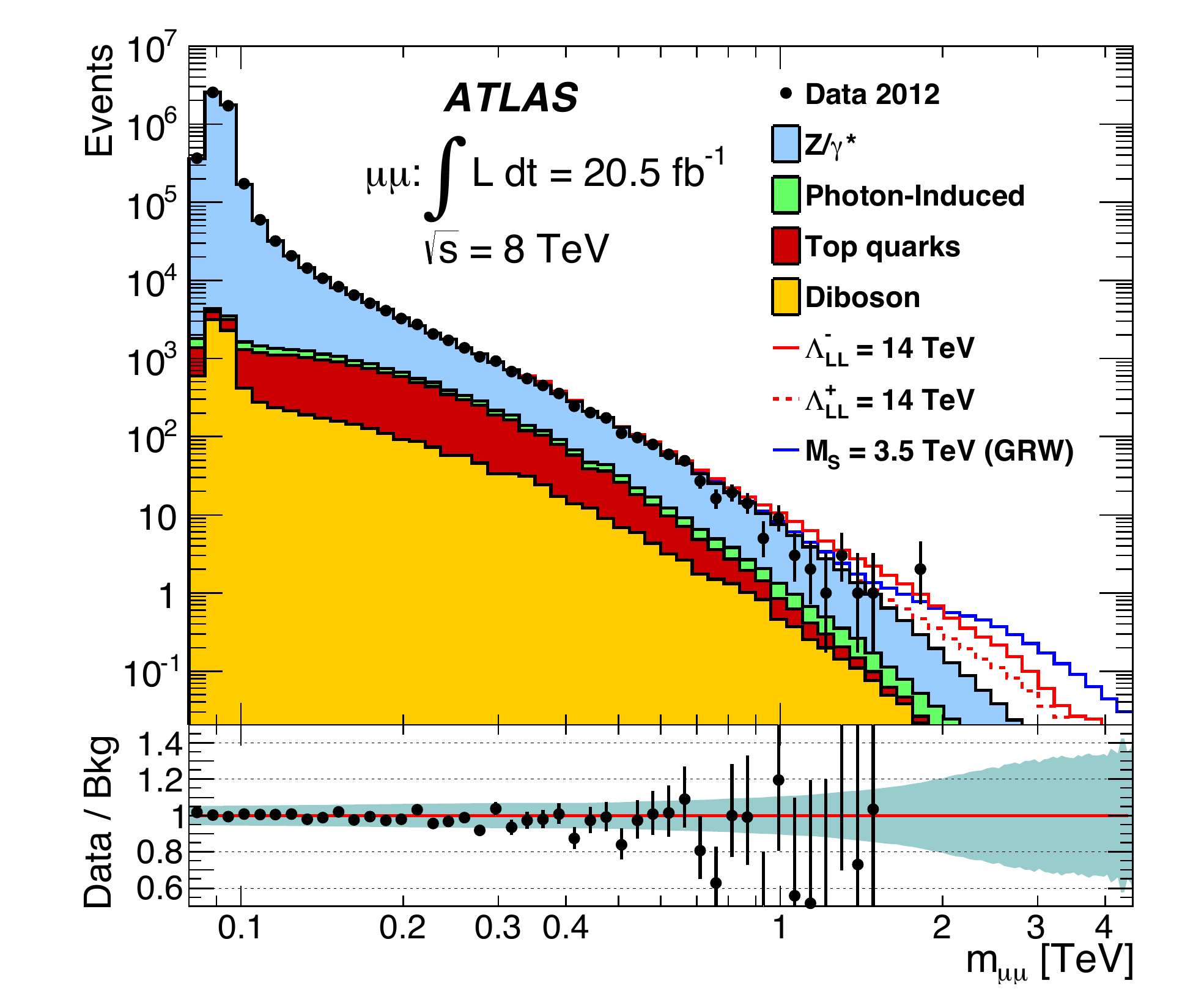}
\includegraphics[width=0.41\columnwidth]{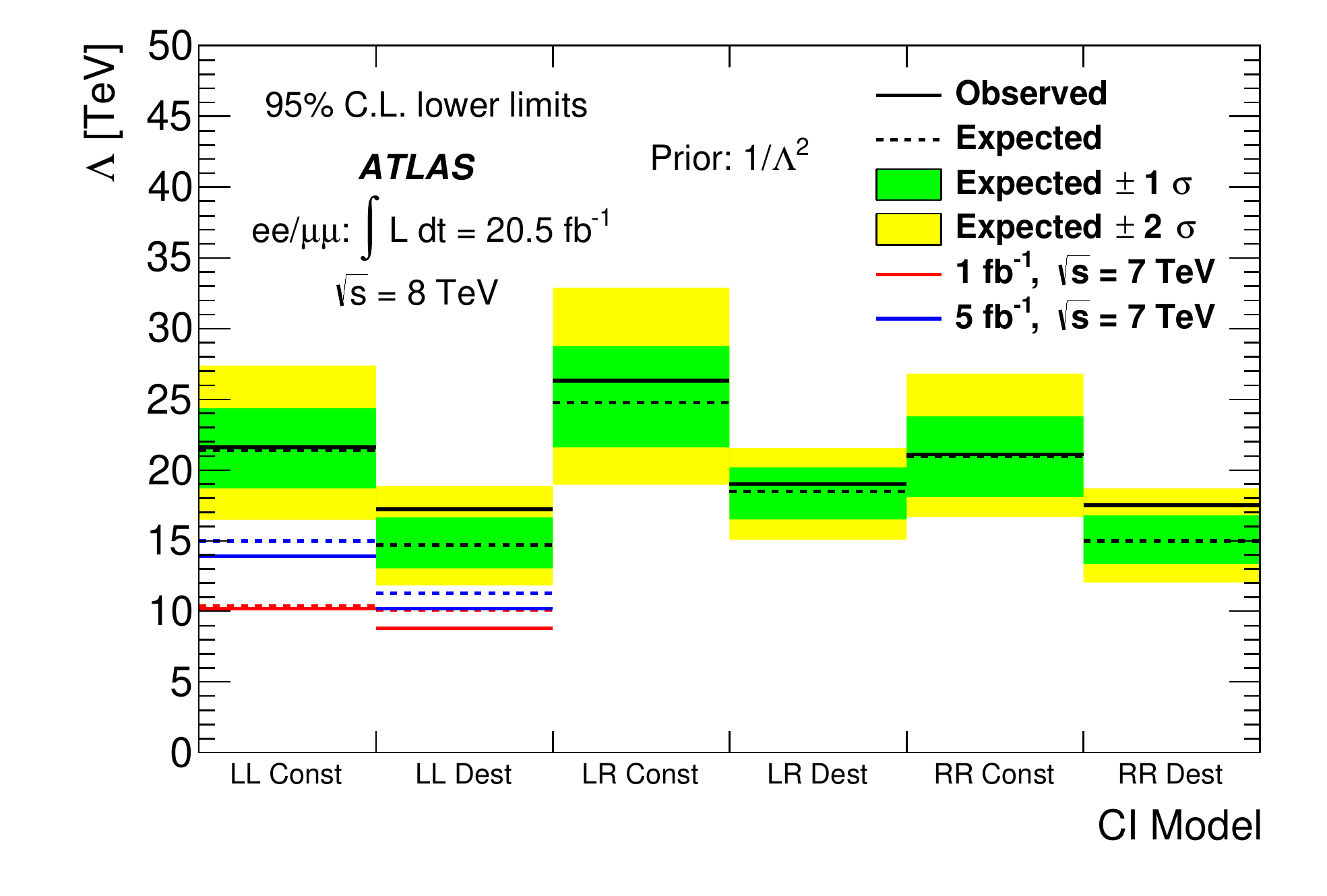}
\caption{Invariant mass of the di-muon system (left) and 95\%CL combined limit on the scale of the di-lepton contact interaction (right). Figures taken from Ref.~\cite{CI}}
\label{fig:plotsCI}
\end{center}
\end{figure}

Instead of searching for a resonance peak on top of the smooth di-lepton mass spectrum, one can also investigate non-resonant phenomena as was done in Ref.~\cite{CI}. This is motivated for example by models of fermion compositeness described by a four-fermion contact interaction~\cite{Eichten:1983hw}. The lepton \pt{} requirements are the same as for the resonance search and leptons have to have opposite charge. The invariant mass is considered starting from 80\,GeV, its distribution is shown in the left panel of figure~\ref{fig:plotsCI} for the muon channel. In contrast to the resonance search (cf. fig.~\ref{fig:plotsZprime} left), a signal is expected to manifest itself not as a distinct peak but rather as an excess over the SM prediction in the tail of the distribution. Since no significant excess is observed, limits on the contact interaction scale $\Lambda$ are derived assuming different chiral structures, leading to destructive or constructive interference with the Drell-Yan process. The results are shown in the right panel of figure~\ref{fig:plotsCI}, the limits are of the order of 15-25\,TeV.
%
\subsection{Extra Dimensions}
\begin{figure}[!ht]
\begin{center}
\includegraphics[width=0.33\columnwidth]{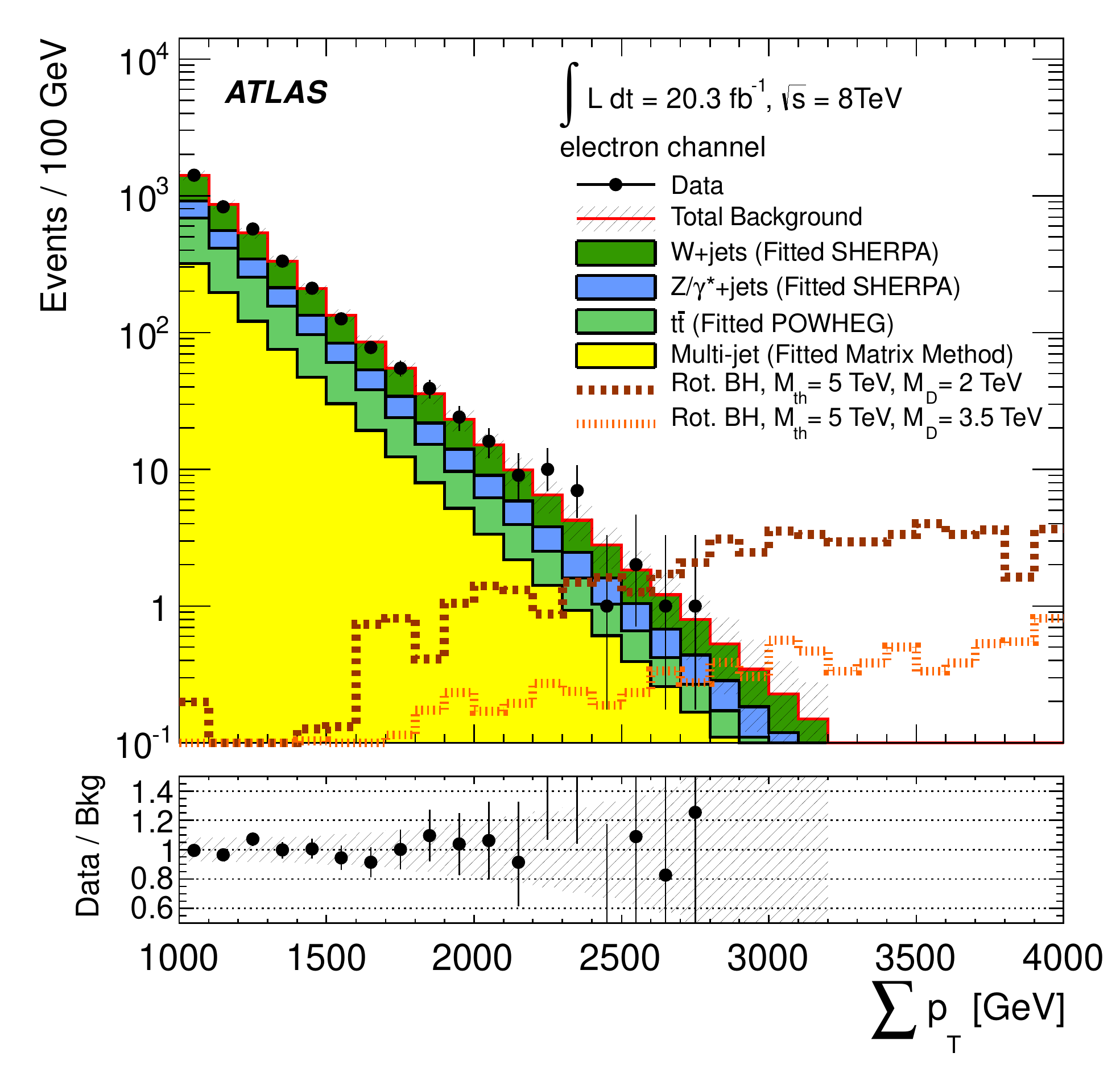}
\includegraphics[width=0.34\columnwidth]{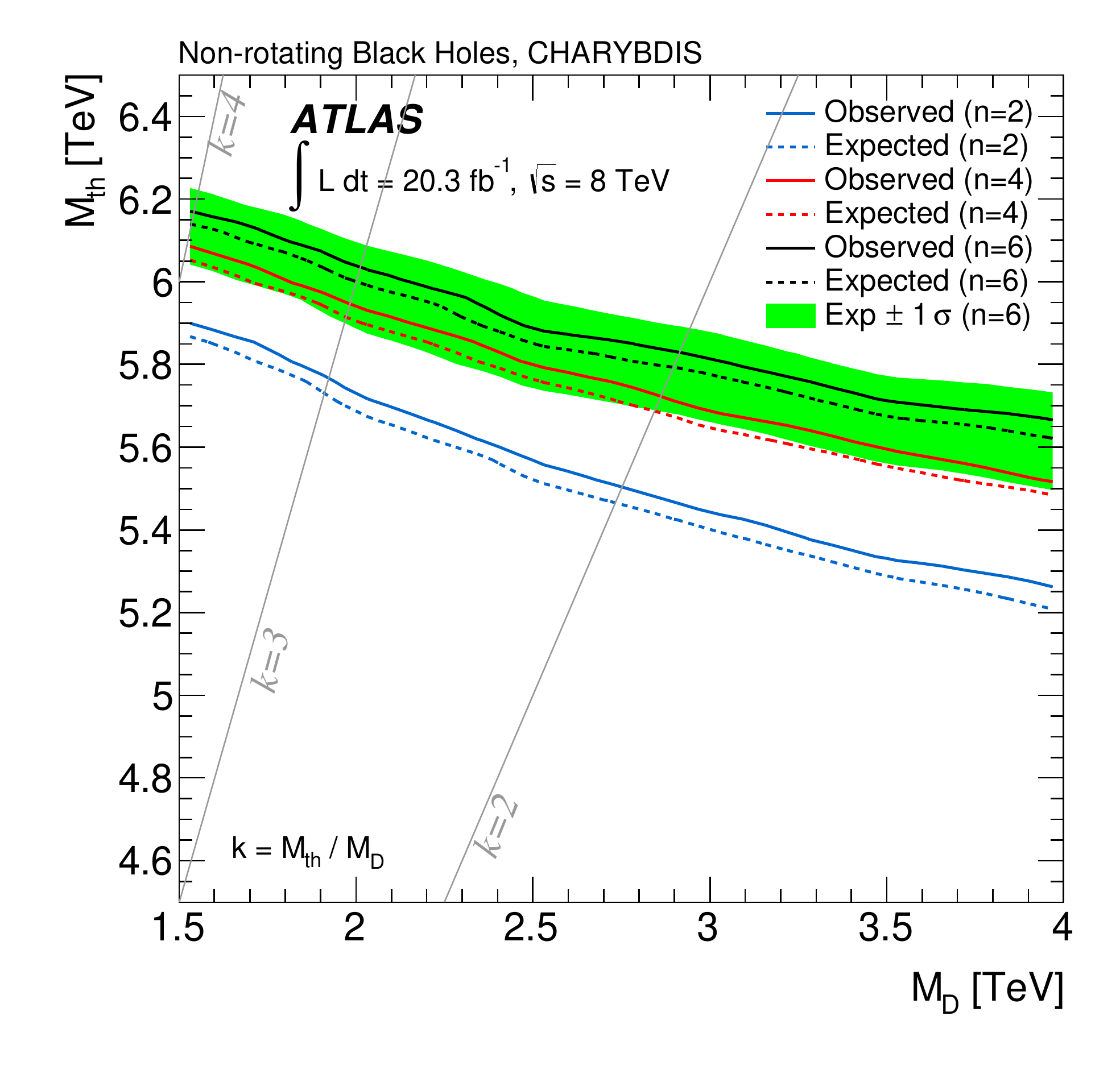}
\caption{Scalar \pt{} sum in the electron channel (left) and combined limit in the $M_{\text{D}}$-$M_{\text{th}}$-plane for different numbers of extra dimensions and the case of non-rotating black holes (right). Figures taken from Ref.~\cite{BH}}
\label{fig:plotsBH}
\end{center}
\end{figure}

A possible solution to the hierarchy problem is provided by models of $n$ extra spacial dimensions of size $R$, as for example the ADD model~\cite{ADD} (for more model references, please consult Ref.~\cite{BH}). The fundamental Planck scale $M_{\text{D}}$ is then related to the usual Planck scale $M_{\text{Pl}}$ via $M_{\text{Pl}}=M_{\text{D}}^{n+2}R^n$. This means, that the fundamental Planck scale could be of the order of TeV, which in turn allows for the production of black holes at LHC energies with some production mass threshold $M_{\text{th}}$. The ATLAS collaboration has used events with at least three objects (electron, muons or jets) with a \pt{} above 100\,GeV, at least one of which has to be a lepton, to search for evidence of black hole production~\cite{BH}. The discriminant variable is the scalar \pt{} sum of the leptons and jets in the event and is shown in the left panel of figure~\ref{fig:plotsBH}. Again, a signal is expected to appear in the tail of the distribution, but no significant excess is found. Limits are derived for different model configurations, one example for non-rotating black holes is shown in the right panel of figure~\ref{fig:plotsBH}. The expected limits are displayed as dashed lines, the observed ones as solid lines. Different numbers of extra dimensions are shown in different colours. For $n=6$, the $\pm1\sigma$ uncertainty band for the expected limit is shown as well.
%
%
\subsection{Di-jet Searches}
\begin{figure}[!ht]
\begin{center}
\includegraphics[width=0.33\columnwidth]{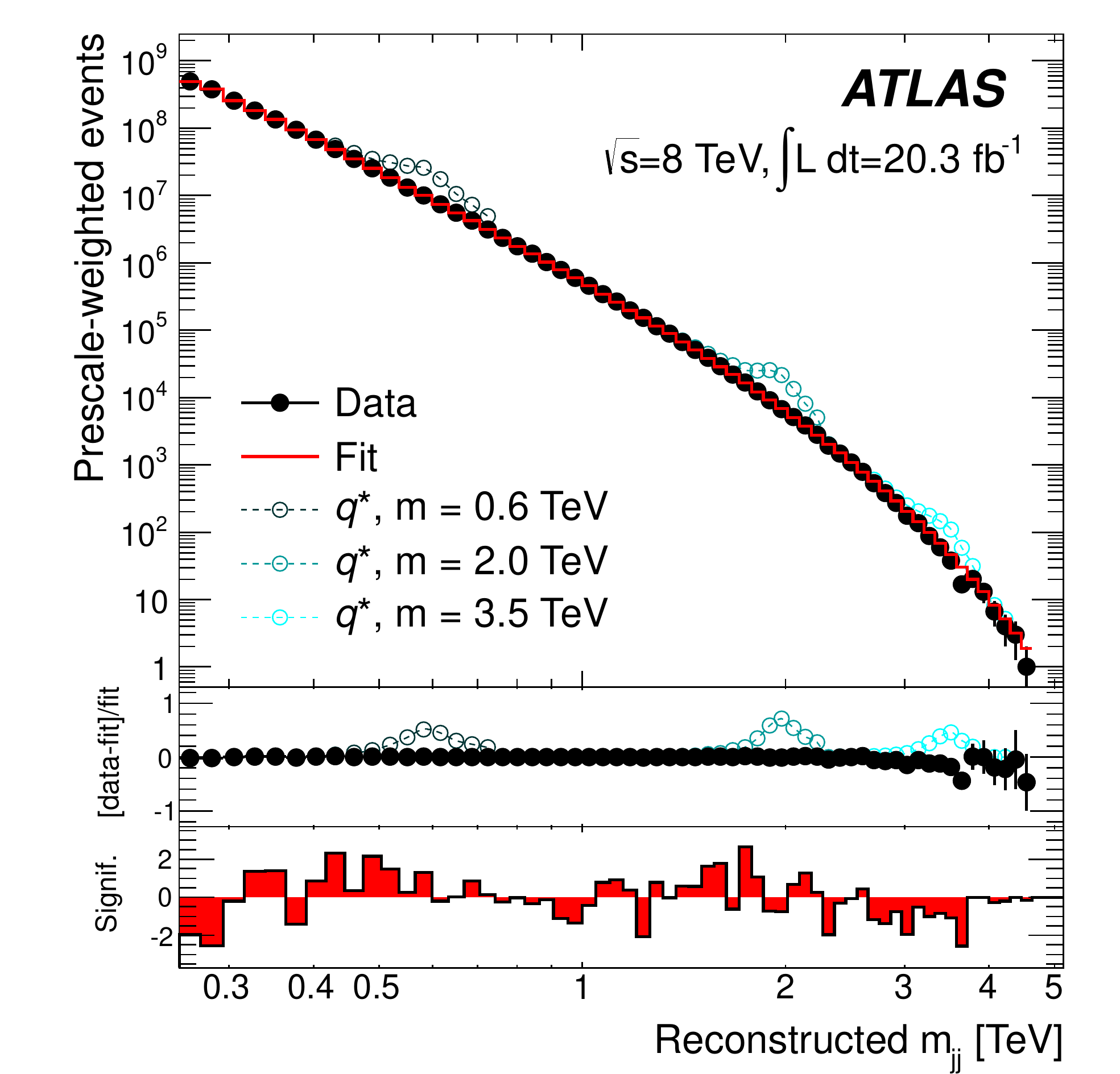}
\includegraphics[width=0.34\columnwidth]{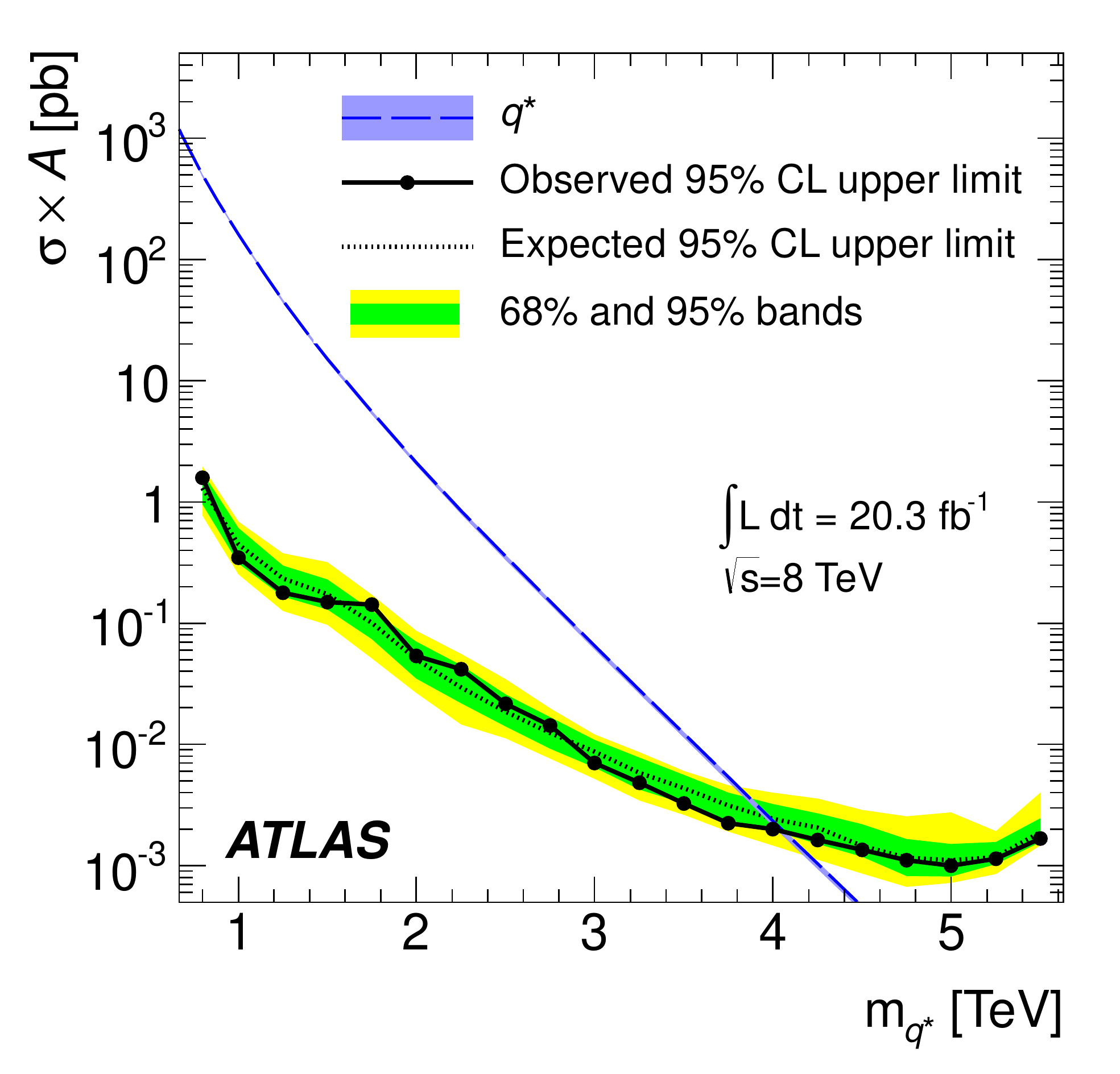}
\caption{Invariant di-jet mass (left) and 95\%CL limit on the visible cross section for the production of excited quarks (right). Figures taken from Ref.~\cite{dijet}}
\label{fig:plotsDijet}
\end{center}
\end{figure}

A very promising signature for the detection of new physics are di-jet events. Quantum chromodynamics (QCD) predicts a smoothly falling spectrum of the invariant mass of the two-jet system, while there are many BSM physics models predicting narrow resonances on top of this, as is shown in the left panel of figure~\ref{fig:plotsDijet}, taken from the ATLAS analysis of di-jet events~\cite{dijet}. The analysis uses events with at least two jets with \pt{}$>50$\,GeV. The absolute rapidity difference of the two jets has to be less than 1.2. As is seen from the $m_{jj}$ distribution, no significant excess is observed and limits on parameters of various different models can be derived. The right panel of figure~\ref{fig:plotsDijet} shows the example of excited quark production. Such models of composite fermions might explain the family structure of the SM, see for example~\cite{Baur:1987ga} or the more complete list in~\cite{dijet}. The limit on the mass of excited quarks is approximately 4\,TeV.
\section{Dark Matter Searches}
The existence of Dark Matter (DM) is firmly established by a large number of cosmological and astrophysical observations, but little is known about what it is made of. A popular class of candidates are Weakly Interacting Massive Particles (WIMPs). 
At a collider, one searches for pair production of WIMPs, which escape the detector without interacting. Therefore, an additional object is needed, typically radiated in the initial state and recoiling against the escaping WIMP pair, leading to missing transverse energy~\cite{Birkedal:2004xn,Beltran:2010ww,Goodman:2010yf,Bai:2010hh,Goodman:2010ku,Carpenter:2012rg,Bai:2012xg,Cotta:2012nj} (for a more exhaustive list please also consult the ATLAS papers referenced below). These signatures are referred to as mono-X, and ATLAS has performed searches for Dark Matter in the mono-jet, mono-photon, mono-$W$ and mono-$Z$ channels. 
All of these searches use an effective field theory (EFT) to describe the interaction between SM and DM particles, which makes the assumption that the interaction is mediated by a particle which is itself too heavy to be directly produced at the LHC and can thus be integrated out, leading to a contact interaction. Different operators describe different initial states and different types of interactions, a subset of all possible operators is used by the different mono-X analyses, cf. for example table 1 in Ref.~\cite{monojet2011}. 
The cross section for the DM pair production via such an operator depends on the mass of the WIMP ($m_{\chi}$) and the suppression scale of the effective theory, $M_{*}$. The virtue of the effective theory is that it allows for a direct comparison to the results of direct searches, which look for the recoil of a target nuclei due to a WIMP scattering off it. The different mono-X analyses will be outlined in the following sections and the results will be summarised and compared in section~\ref{subsec:DMResSum}.
\subsection{Mono-Jet}
Initial state radiation (ISR) of a jet has a large cross section at the LHC and therefore the mono-jet analysis is a promising channel for the DM search. It is moreover the only of the above mentioned analyses that is sensitive to the gluon-gluon operator (labelled D11, cf. Ref.~\cite{monojet2011}). Events are required to have one highly energetic jet with a \pt{} greater than 120\,GeV. One further jet is allowed in the event to account for additional radiation or splitting. Events with electrons or muons are rejected. The search is performed in 4 signal regions (SRs) of increasing cuts on the leading jet \pt{} and \met{}, with values of 120, 220, 350 and 500\,GeV. The \met{} spectrum obtained with 10\,\ifb{} of $\sqrt{s}=8$\,TeV data is presented on the left in figure~\ref{fig:metJetWZhadPlots} in comparison to the SM prediction and possible signal distributions. The data is in agreement with the background-only hypothesis. 
\begin{figure}[!ht]
\begin{center}
\includegraphics[width=0.4\columnwidth]{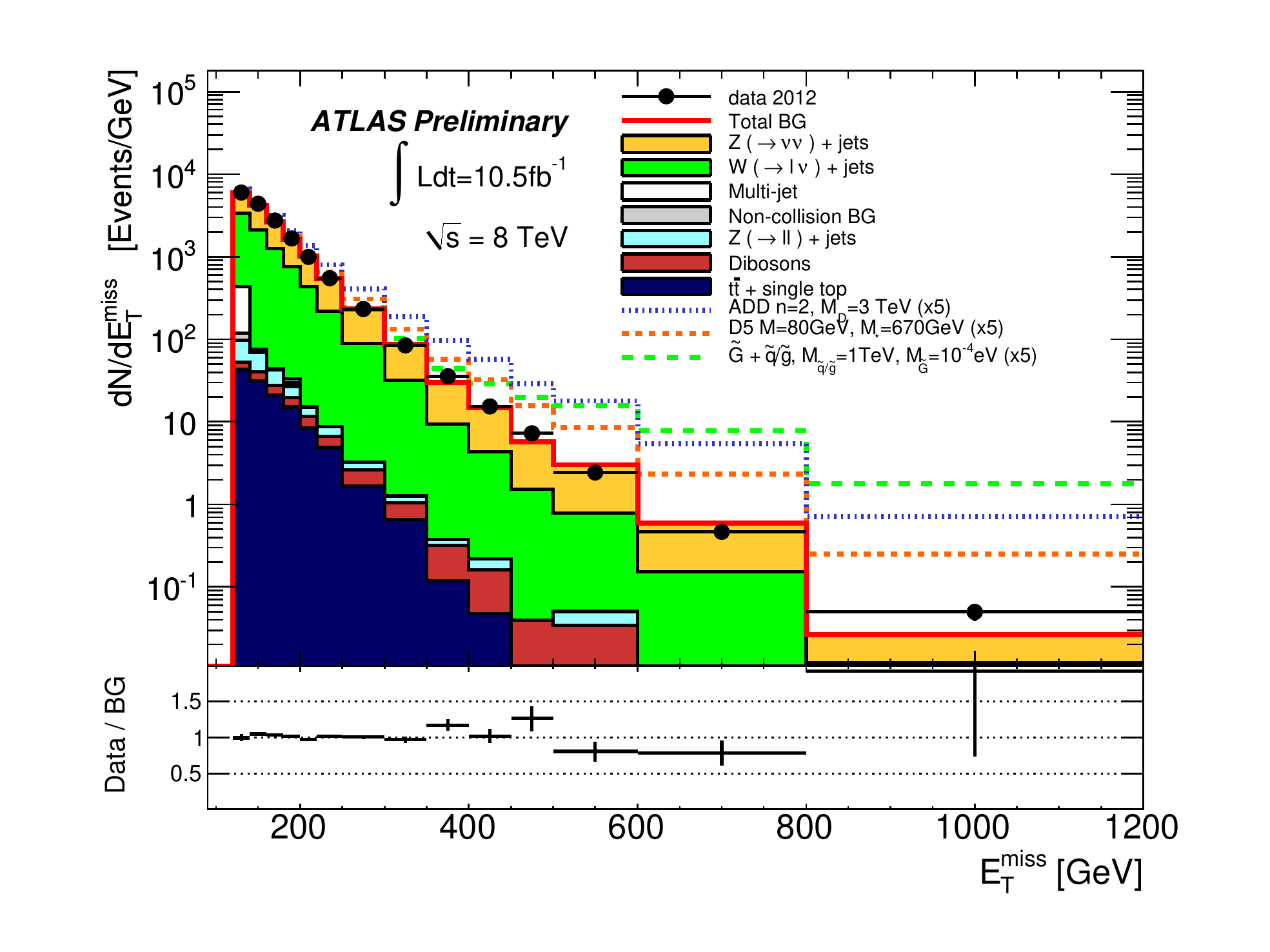}
\includegraphics[width=0.43\columnwidth]{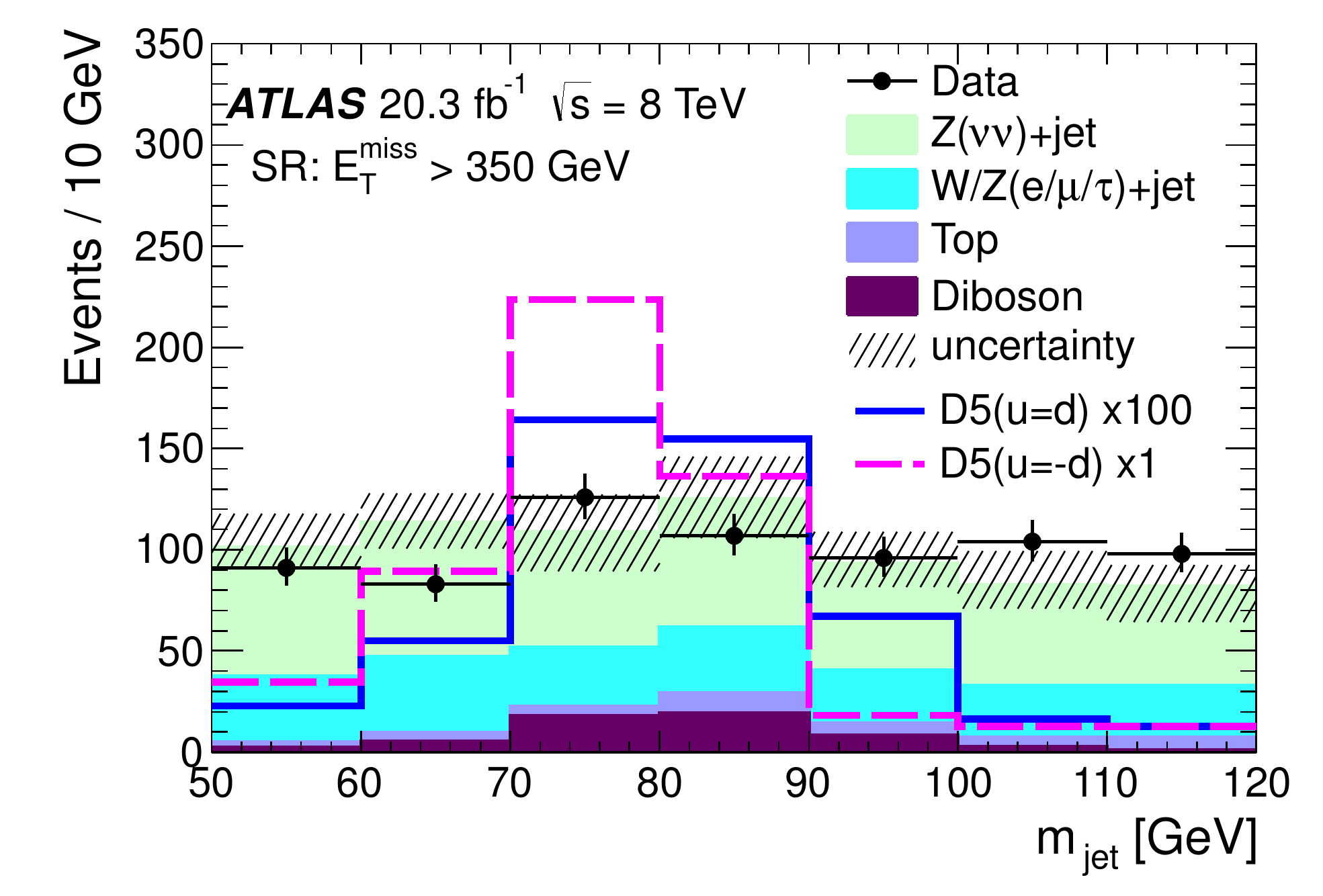}
\caption{Left: \met{} distribution in the first signal region from the mono-jet analysis~\cite{monojetHCP}, right: mass of the fat jet in the hadronic mono-$W/Z$ analysis~\cite{monoWZhad}.}
\label{fig:metJetWZhadPlots}
\end{center}
\end{figure}
\subsection{Mono-Photon}
Events with a photon radiated in the initial state from the complete $\sqrt{s}=7$\,TeV data set have been used to search for Dark Matter candidates as well~\cite{monophoton}. The photon is required to have a \pt{} exceeding 150\,GeV. Electrons and muons are vetoed, and only one jet is allowed in the events. The \met{} has to be larger than 150\,GeV as well. 
No excess at large \met{} is observed and limits on the suppression scale and the WIMP-nucleon scattering cross section are derived. These limits are in general weaker than the mono-jet limits due to the smaller cross section.
\subsection{Hadronic Mono-$W$/$Z$}
ATLAS has performed a search for Dark Matter also using events with hadronically decaying $W$ or $Z$ bosons and missing transverse energy~\cite{monoWZhad}. In this topology, the quark pair from the boson decay is boosted and the resulting jets merge into one object, called a fat jet. The event selection requires one such jet with a \pt{}$>150$\,GeV and \met{} greater than 350 or 500\,GeV. Events are vetoed if they contain a lepton, a photon or more than one additional jet. The mass of the fat jet is the discriminant variable, shown in the right panel of figure~\ref{fig:metJetWZhadPlots}. Once more, no significant excess is observed.
\subsection{Leptonic Mono-$W$/$Z$}
\subsubsection{Mono-$Z(\ell^{+}\ell^{-})$}
The leptonic decay of $Z$-bosons in events with large \met{} is also used to search for Dark Matter production~\cite{monoZlep}. In this case, there is an additional Feynman diagram where the contact interaction is between the DM and the $Z$ itself. Events are selected if they contain two oppositely charged electrons or muons with additional requirements on the di-lepton system, and events with additional leptons or jets are vetoed. Four signal regions are considered with \met{} cuts at 150, 250, 350, and 450\,GeV. The \met{} distribution in the lowest SR is shown on the left of figure~\ref{fig:mWZlep}: Again, there is no excess observed in data.
\begin{figure}[!ht]
\begin{center}
\includegraphics[width=0.37\columnwidth]{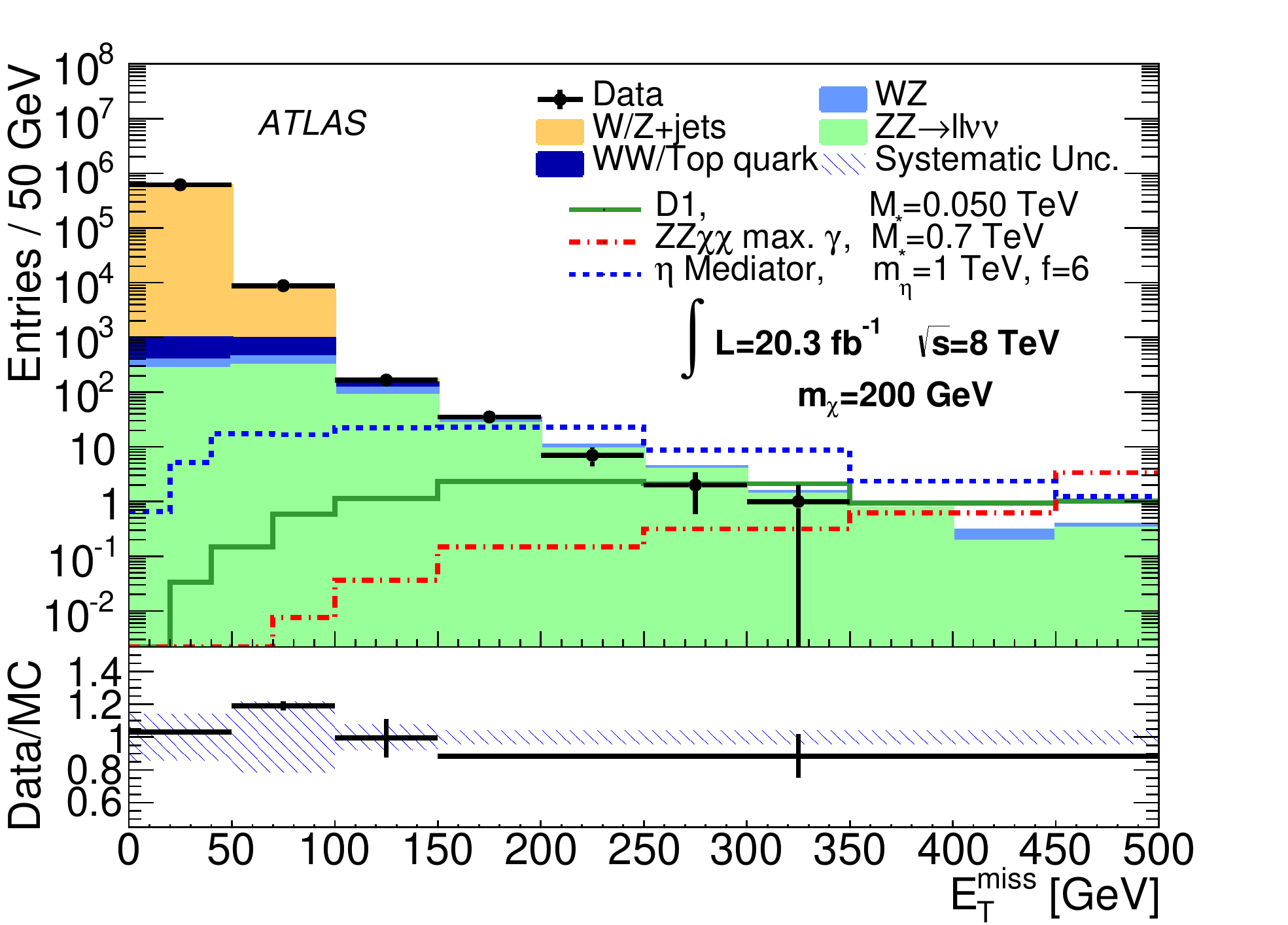}
\includegraphics[width=0.38\columnwidth]{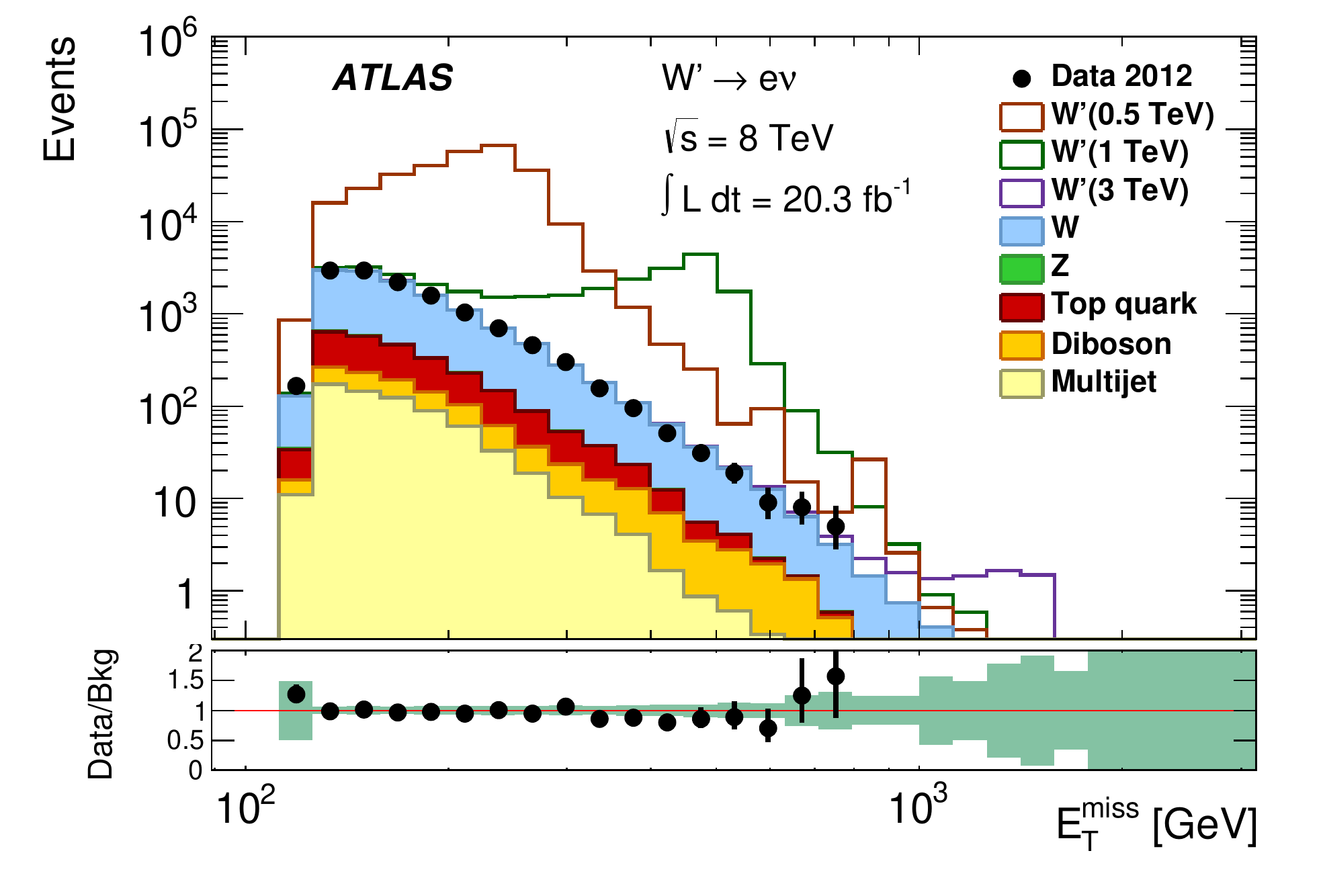}
\caption{\met{} distribution obtained in the leptonic mono-$Z$ analysis~\cite{monoZlep} (left) and the leptonic mono-$W$ analysis~\cite{Wprime} (right).}
\label{fig:mWZlep}
\end{center}
\end{figure}%
\subsubsection{Mono-$W(\ell^{\pm}\parenbar{\nu})$}
The leptonic mono-$W$ Dark Matter search uses the same event selection as the search for new heavy gauge bosons~\cite{Wprime}, already presented in section~\ref{subsubsec:lepMet}. However, the discriminant variable in this case is the missing transverse energy, which is presented on the right in figure~\ref{fig:mWZlep}. It can be seen that no significant excess is observed.
\subsection{Mono-X Results Summary}
\label{subsec:DMResSum}
On the left-hand side of figure~\ref{fig:XN}, taken from Ref.~\cite{Wprime}, limits for spin-independent and spin-dependent interactions from various mono-X and direct detection searches are summarised. For the spin-independent interactions, results obtained for the vector operator (D5) are used, in case of spin-dependent interactions, the limits for the tensor operator (D9) are compared. In the latter case, the collider limits surpass the direct detection limits over almost the entire range of WIMP masses considered. The strongest limits are obtained from the hadronic mono-$W/Z$ analysis, the leptonic mono-$Z$ analysis provides stronger bounds than the leptonic mono-$W$ search. Despite the fact that they were obtained using the much smaller $\sqrt{s}=7$\,TeV data set, the mono-jet limits are almost compatible with the mono-$W$(lep) results.
\begin{figure}[!ht]
\begin{center}
\includegraphics[width=0.55\columnwidth]{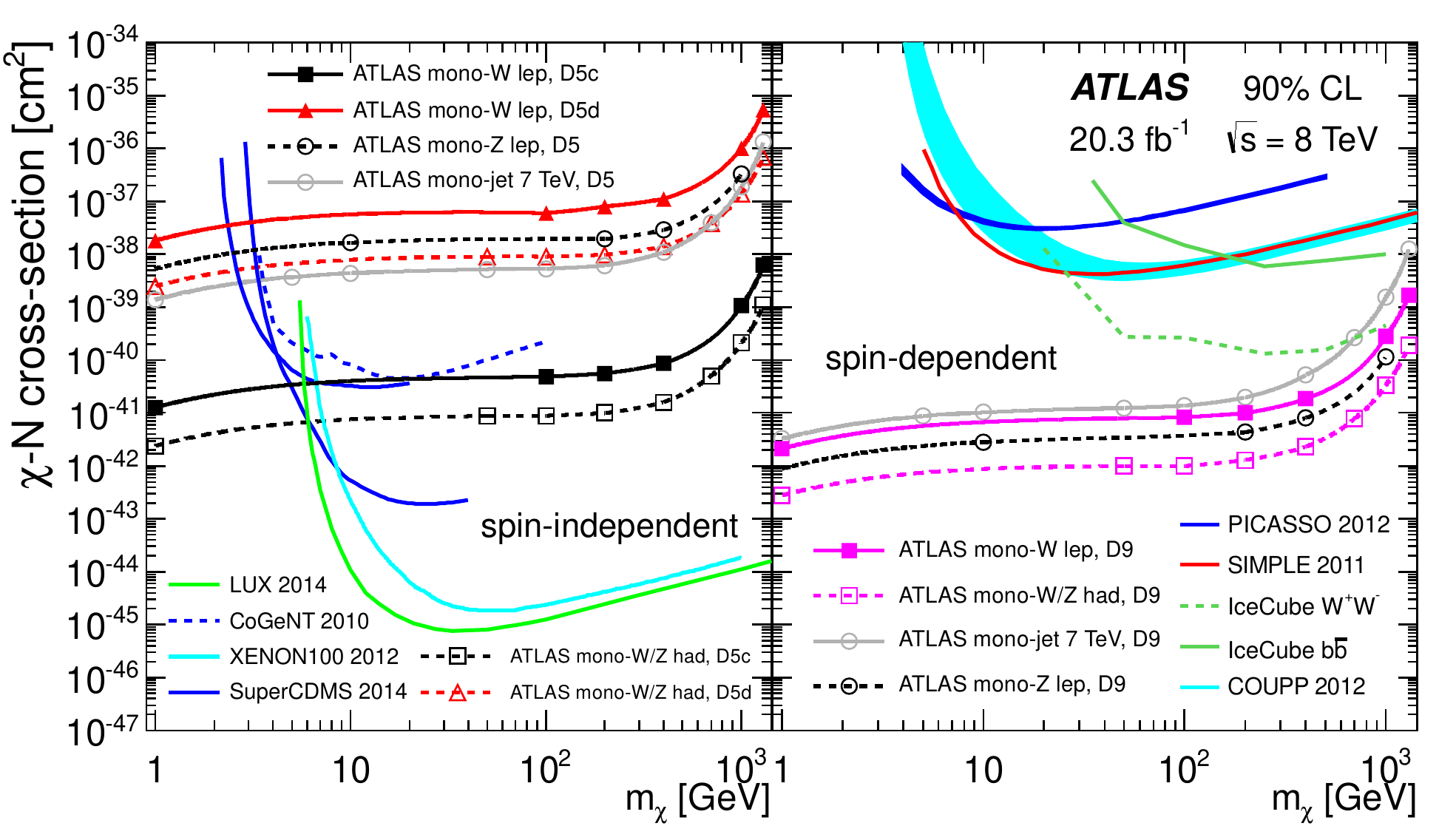}
\includegraphics[width=0.44\columnwidth]{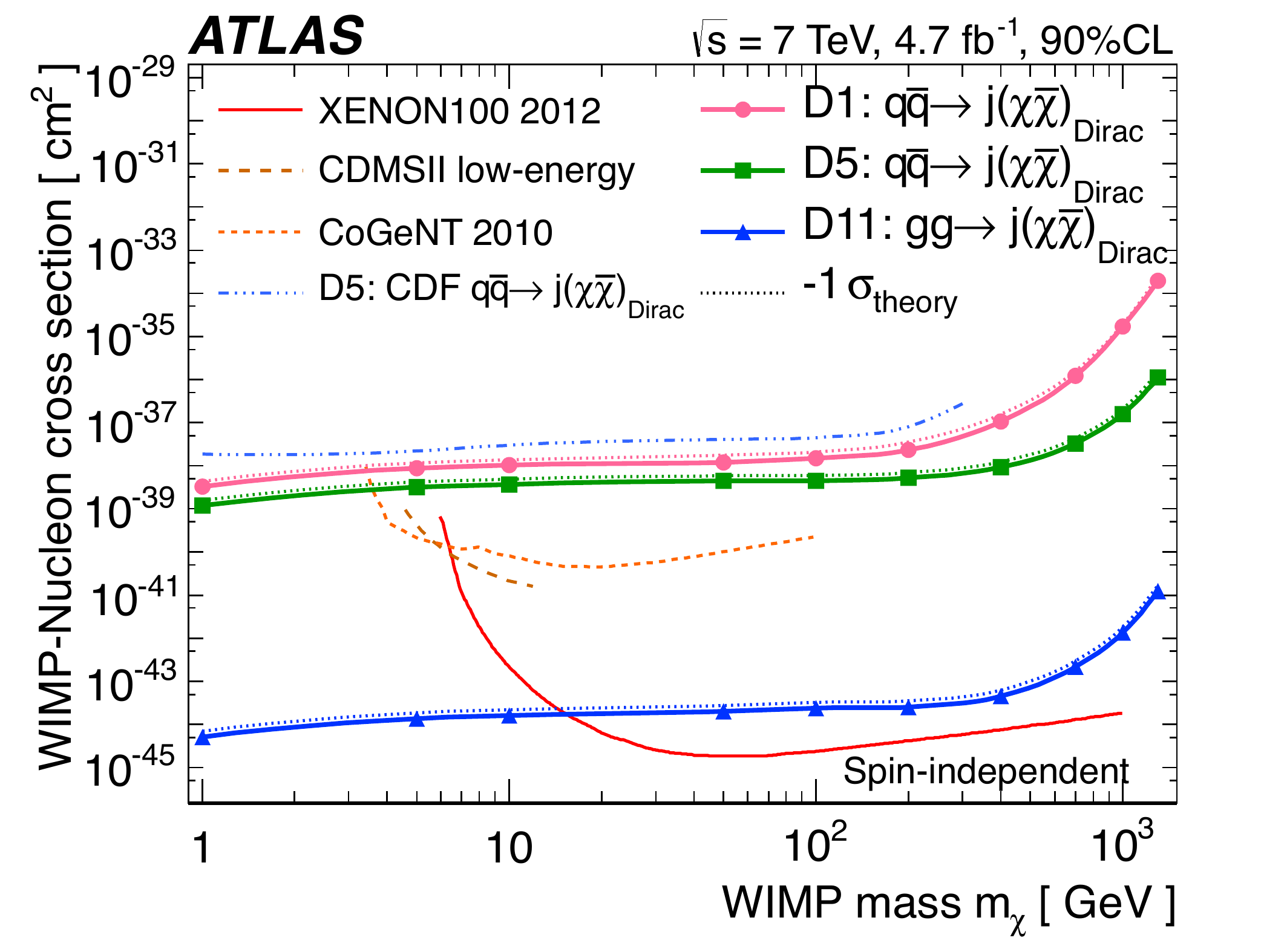}
\caption{90\%CL limits on WIMP-nucleon scattering cross section from various mono-X channels (left)~\cite{Wprime} and from the mono-jet 7\,TeV analysis (right)~\cite{monojet2011}.}
\label{fig:XN}
\end{center}
\end{figure}\\
%
For the operator D5, there is one subtlety for the mono-$W$ analyses: Constructive interference ('D5c') occurs if the couplings to up- and down-quarks have opposite signs but are equally large (in the case of same signs, the interference is destructive ('D5d')). Assuming constructive interference, the mono-$W$ limits surpass the collider limits from other channels by several orders of magnitude, the hadronic analysis again provides the strongest bounds. Among the other channels, the mono-jet limits are strongest, even though they correspond to the $\sqrt{s}=7$\,TeV data set. In comparison to the direct detection limits, the collider bounds are competitive only at small WIMP masses, where the direct searches have no sensitivity. The strongest collider limits for spin-independent interactions are obtained for the gluon-gluon operator D11 that is only probed by the mono-jet analysis. The $\sqrt{s}=7$\,TeV limits are shown on the right of figure~\ref{fig:XN}. It can be seen that for D11 the limits are almost six orders of magnitude stronger than those for the other operators.
\subsection{Simplified Models}
Instead of working with an EFT, the next natural step towards an ultra-violet complete model is a simplified model in which the mediator is not integrated out~\cite{Fox:2011pm}. In Ref.~\cite{monojet14tev}, such a simplified model with an s-channel vector mediator was investigated in the mono-jet analysis. The left plot in figure~\ref{fig:SM} shows the lower limits obtained on the scale ${M_{*}\equiv M_{Med}/\sqrt{g_{SM}\,g_{DM}}}$ as a function of the mediator mass $M_{Med}$, where $g_{SM}$ and $g_{DM}$ are the couplings of the mediator to SM and DM particles, respectively. Blue lines correspond to a WIMP mass of 50\,GeV, orange to 400\,GeV. The different line styles represent different choices of the width of the mediator. Three regimes can be distinguished: at low mediator masses, the mediator has to be produced off-shell leading to small cross sections and weak limits. At intermediate masses a resonant-like enhancement is observed as the mediator can now be produced on-shell. For high mediator masses, the limits approach those obtained in the EFT, as is to be expected.\\
\begin{figure}[!ht]
\begin{center}
\includegraphics[width=0.35\columnwidth]{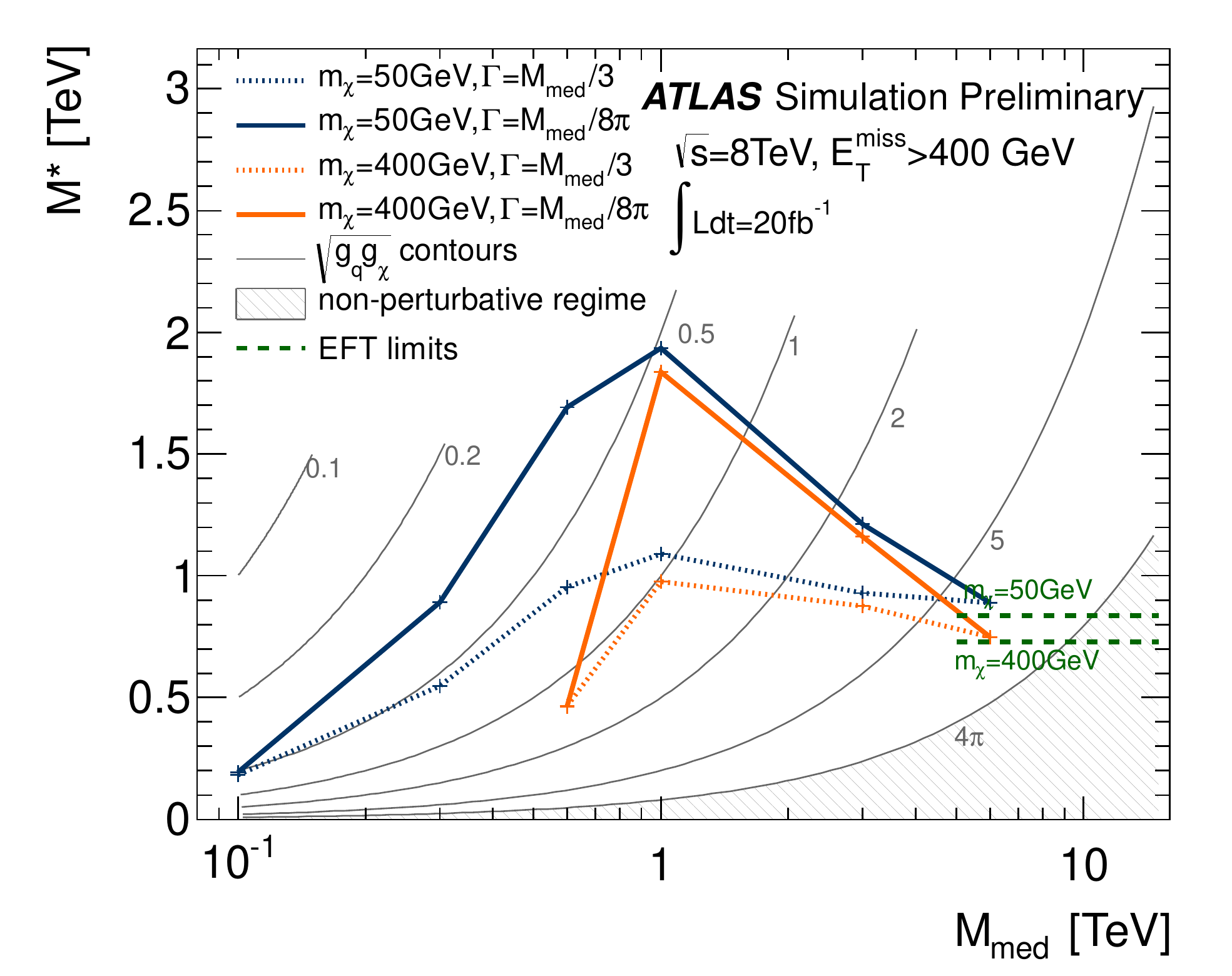}
\includegraphics[width=0.39\columnwidth]{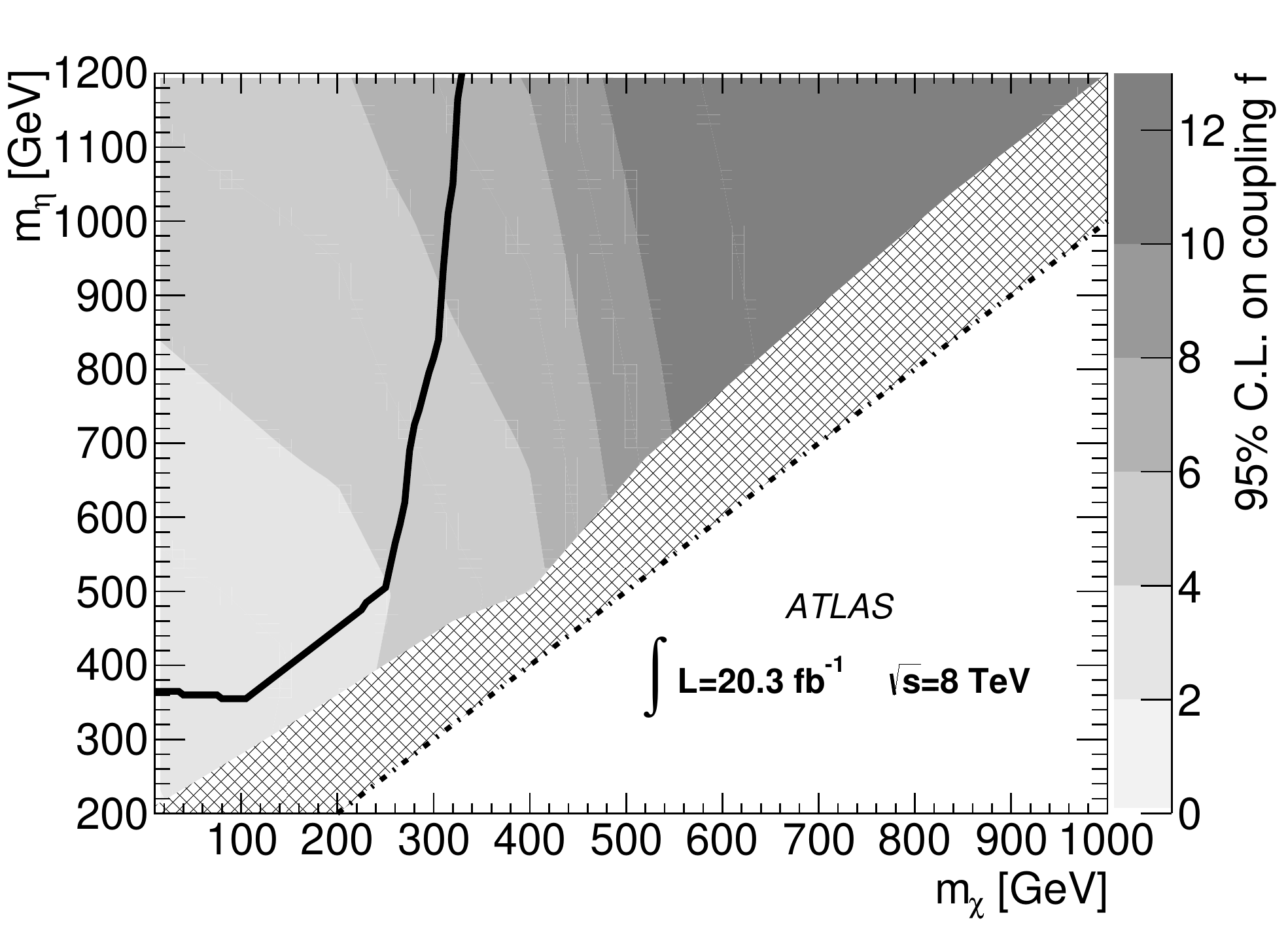}
\caption{ 
95\%CL lower limits on $M_{*}$ as a function of the mediator mass (left)~\cite{monojet14tev} and upper limits on the coupling strength in the plane of WIMP mass and mediator mass (right)~\cite{monoZlep}.}
\label{fig:SM}
\end{center}
\end{figure}%
%
The leptonic mono-$Z$ analysis~\cite{monoZlep} also investigates a simplified model with a scalar, coloured t-channel mediator (labelled $\eta$). Upper limits on the couplings ($f$) are derived in the plane of mediator mass and WIMP mass, as shown on the right in figure~\ref{fig:SM}. In the region top-left of the black line, the limits are in conflict with the observed relic abundance of Dark Matter.
\subsection{Future Prospects}
Prospects for the mono-jet and di-lepton analysis at $\sqrt{s}=14$\,TeV have been studied~\cite{monojet14tev, dilep14tev}. \\
The expected limit on the $Z'$ mass for 3000\,\ifb{} at $\sqrt{s}=14$\,TeV is of the order of 8\,TeV, while the current $\sqrt{s}=8$\,TeV limit is approximately 3\,TeV.\\
The expectations for the limits for the effective operator D5 in case of the mono-jet analysis are shown in figure~\ref{fig:mj14}. The left plot shows the expected limits as function of the \met{} cut for both 8\,TeV (orange) and 14\,TeV (blue) for two different WIMP masses (50\,GeV, 400\,GeV) and for luminosities corresponding to approximately one year of data taking each. It is found that the limits improve by approximately a factor of two and that the gain is largest in regions of high \met{}. The plot on the right of figure~\ref{fig:mj14} shows the evolution of the expected limits at 14\,TeV with increasing luminosity. This shows that more luminosity only increases the sensitivity up to a certain point and that a further gain can only be achieved by going to regions of higher \met{}.
\begin{figure}[!ht]
\begin{center}
\includegraphics[width=0.38\columnwidth]{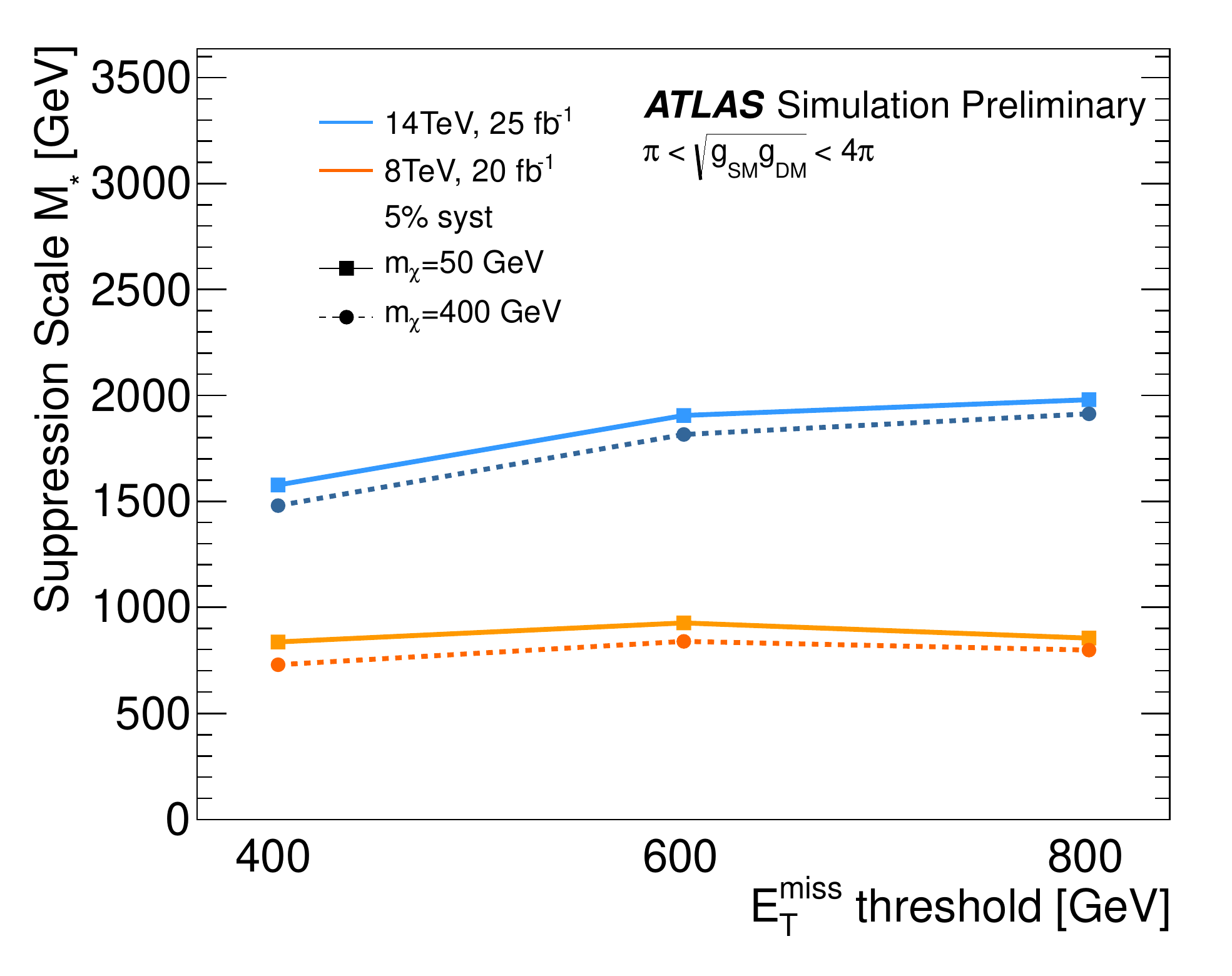}
\includegraphics[width=0.38\columnwidth]{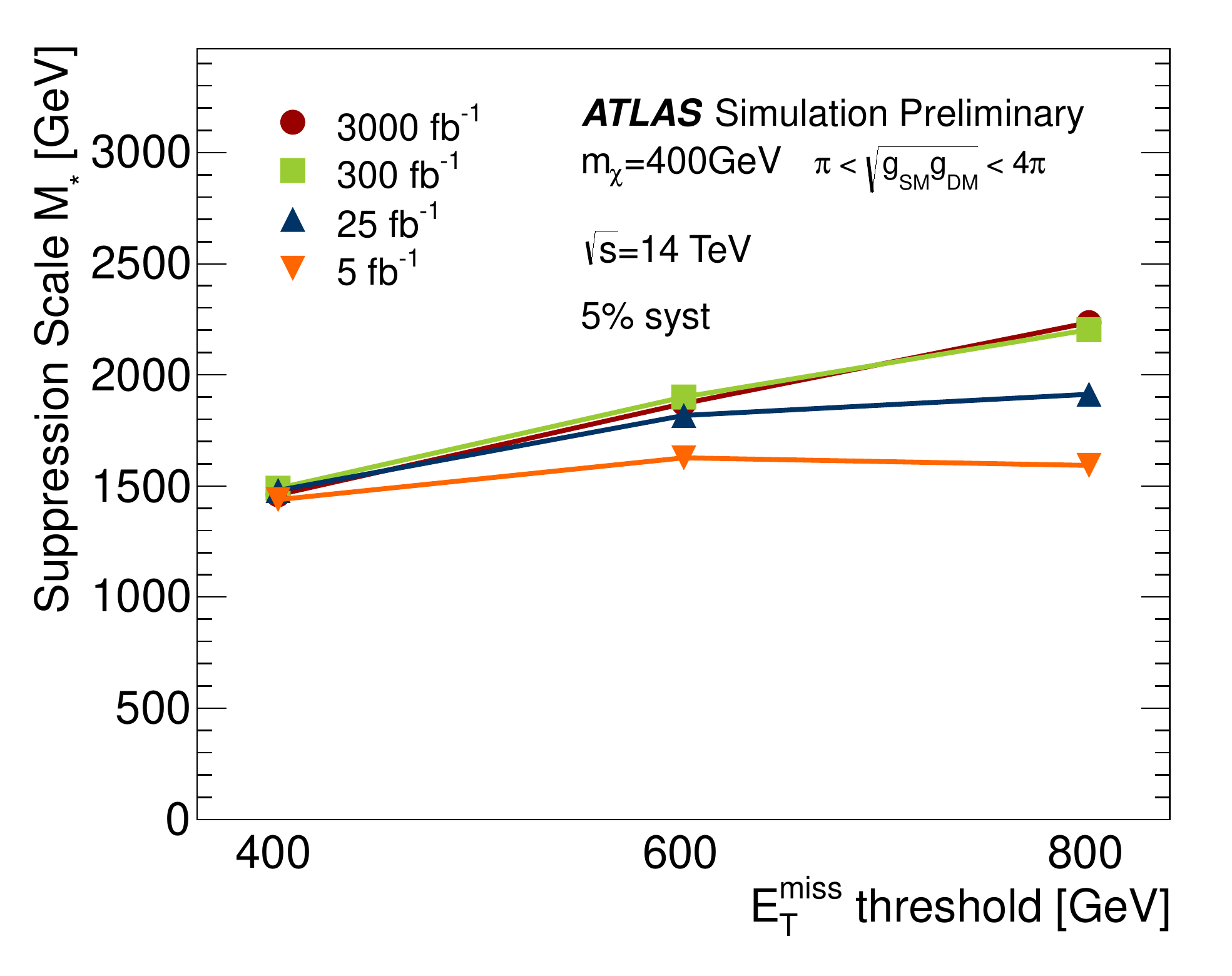}
\caption{Expected limits on $M_{*}$ for 8 and 14\,TeV (left) and evolution of expected limits for different luminosities at 14\,TeV (right). Figures taken from Ref.~\cite{monojet14tev}}
\label{fig:mj14}
\end{center}
\end{figure}%
\section{Summary}
The ATLAS collaboration has performed searches for numerous scenarios of BSM physics in many different final states, using the data set delivered during the first run of the LHC. No significant excess over the Standard Model expectation is observed and limits are derived on various model parameters. Limits on the masses of new heavy Gauge bosons are on the order of 2.5-3\,TeV, limits on the mass of black holes are on the order of 5-6\,TeV. The scale of di-lepton contact interactions is bounded from below to be on the order of 20\,TeV, limits on excited-quark masses reach approximately 4\,TeV. \\
Searches for Dark Matter are conducted in many mono-X channels, each of which have their own strength. For spin-independent interactions, the collider limits are found to surpass the direct search results at low WIMP masses, in the case of spin-dependent interactions the collider limits are competitive over a wide range of WIMP masses. In addition to the effective field theory, mono-X searches also begin to consider more UV complete, simplified models. Simulation studies for $\sqrt{s}=14$\,TeV indicate that the sensitivity in terms of expected limits can be improved by approximately a factor of two or more.
\bigskip
%

%
%

%
%
%
%
 
\end{document}